\documentclass[10pt,journal,compsoc]{IEEEtran}
\usepackage{hyperref}
\hypersetup{colorlinks=true, citecolor=black}
\usepackage{amsmath,amsfonts,bm}
\usepackage{amssymb}
\usepackage{dcolumn}
\usepackage[utf8]{inputenc}
\usepackage[T1]{fontenc}
\usepackage{ragged2e}
\usepackage{braket}
\usepackage{array}
\usepackage[caption=false,font=normalsize,labelfont=sf,textfont=sf]{subfig}
\usepackage{textcomp}
\usepackage{stfloats}
\usepackage{url}
\usepackage{verbatim}
\usepackage{graphicx}
\usepackage{array, tabularx, booktabs, lipsum, multirow}
\usepackage{xcolor}
\usepackage{siunitx}
\usepackage{tikz}
\hypersetup{colorlinks=true, citecolor=black, urlcolor=black, linkcolor=black}
\usepackage[qm]{qcircuit}
\usepackage{balance}
\usepackage{titlesec}

\ifCLASSOPTIONcompsoc
  \usepackage[nocompress]{cite}
\else
  \usepackage{cite}
\fi

\ifCLASSINFOpdf
\else
\fi
\newcommand{\figref}[2]{\hyperref[#1]{\ref*{#1}#2}}

\begin{document}

\title{Noise-Resistant Feature-Aware Attack Detection Using Quantum Machine Learning}

\author{Chao Ding, Shi Wang, Jingtao Sun, Yaonan Wang, Daoyi Dong, and Weibo Gao~
\IEEEcompsocitemizethanks{
\IEEEcompsocthanksitem Chao Ding is with the School of Artificial Intelligence and Robotics, Hunan University, Changsha 410082, China, and also with the Centre for Quantum Technologies, National University of Singapore, Singapore 117543, Singapore (e-mail: dingchao5170@hnu.edu.cn).\protect \vspace{0.2ex}
\IEEEcompsocthanksitem Shi Wang, Jingtao Sun, and Yaonan Wang are with the School of Artificial Intelligence and Robotics, Hunan University, Changsha 410082, China (e-mail: shi{\_}wang@hnu.edu.cn; jingtaosun@hnu.edu.cn; yaonan@hnu.edu.cn).\protect \vspace{0.2ex}
\IEEEcompsocthanksitem Daoyi Dong is with the Australian Artificial Intelligence Institute, Faculty of Engineering and Information Technology, University of Technology Sydney, NSW 2007, Australia (e-mail: daoyidong@gmail.com).\protect \vspace{0.2ex}
\IEEEcompsocthanksitem Weibo Gao is with the School of Electrical and Electronic Engineering, Nanyang Technological University, Singapore 639798, Singapore; the Division of Physics and Applied Physics, School of Physical and Mathematical Sciences, Nanyang Technological University, Singapore 637371, Singapore; the Centre for Quantum Technologies, National University of Singapore, Singapore 117543, Singapore; and the Quantum Science and Engineering Centre (QSec), Nanyang Technological University, Singapore 639798, Singapore (e-mail: wbgao@ntu.edu.sg).\protect \vspace{1.5ex}
}
\thanks{Manuscript received [redacted]; revised [redacted].}
}

\IEEEtitleabstractindextext{
\begin{justify}
	\begin{abstract}
		Continuous-variable quantum key distribution (CV-QKD) is a quantum communication technology that offers an unconditional security guarantee. However, the practical deployment of CV-QKD systems remains vulnerable to various quantum attacks. In this paper, we propose a quantum machine learning (QML)-based attack detection framework (QML-ADF) that safeguards the security of high-rate CV-QKD systems. In particular, two alternative QML models—quantum support vector machines (QSVM) and quantum neural networks (QNN)—are developed to perform noise-resistant and feature-aware attack detection before conventional data postprocessing. Leveraging feature-rich quantum data from Gaussian modulation and homodyne detection, the QML-ADF effectively detects quantum attacks, including both known and unknown types defined by these distinctive features. The results indicate that all twelve distinct QML variants for both QSVM and QNN exhibit remarkable performance in detecting both known and previously undiscovered quantum attacks, with the best-performing QSVM variant outperforming the top QNN counterpart. Furthermore, we systematically evaluate the performance of the QML-ADF under various physically interpretable noise backends, demonstrating its strong robustness and superior detection performance. We anticipate that the QML-ADF will not only enable robust detection of quantum attacks under realistic deployment conditions but also strengthen the practical security of quantum communication systems.
	\end{abstract}
\end{justify}

\begin{IEEEkeywords}
\justifying
Quantum machine learning, continuous-variable quantum key distribution, quantum information processing, quantum attacks, quantum circuits, feature extraction, attack detection, noise modeling.
\end{IEEEkeywords}}

\maketitle

\IEEEdisplaynontitleabstractindextext

%
\IEEEpeerreviewmaketitle

\IEEEraisesectionheading{\section{introduction}\label{introduction}}
\IEEEPARstart{Q}{uantum} key distribution~(QKD) \cite{cao2022evolution, li2025decentralized}, particularly continuous-variable~QKD~(CV-QKD),~is~a~promising candidate for large-scale and secure quantum communication due to its compatibility with standard optical telecommunication technologies and its potential for high secret key rates \cite{weedbrook2012gaussian, Jouguet-2013, huang2016long, Diamanti-2016, ma2018phase, Xu-2020, zhang2023ic, mehic2020quantum}. Theoretically, CV-QKD has been demonstrated to be unconditionally secure in both the asymptotic \cite{garcia2006unconditional} and finite-size regimes \cite{leverrier2010finite}. However, the deployment of CV-QKD systems in practice is vulnerable to various quantum attacks \cite{jain2016attacks}, which may exploit imperfections in physical devices or constraints in operational procedures, severely compromising system security. For example, there are several typical quantum attacks, including homodyne-detector-blinding attacks \cite{Qin-2018}, local oscillator (LO)-intensity attacks \cite{ma-2013}, calibration attacks \cite{Jouguet-2013-Attack}, saturation attacks \cite{Qin-2016}, and wavelength attacks \cite{huang-2014}.

Fortunately, a number of effective countermeasures \cite{Jouguet-2013-Attack, Ma-2014, Liu-2017, Mao-2020, mao2020hidden, liao2022, Luo-2022, ding2023machine, kish2024mitigation} have been developed to address the challenges posed by quantum attacks. These countermeasures typically fall into two categories: deploying real-time monitoring modules \cite{Jouguet-2013-Attack, Ma-2014, Liu-2017} and designing machine learning-based attack detection frameworks \cite{Mao-2020, mao2020hidden, liao2022, Luo-2022, ding2023machine, kish2024mitigation}. The former approach strengthens the practical security of CV-QKD systems by continuously monitoring critical physical parameters of optical pulses—such as phase, wavelength, and intensity—to counteract quantum attacks. However, these monitoring modules are typically designed to detect specific types of attacks, making them inadequate for handling the potential threats. Moreover, the quantum attacks initiated by an eavesdropper (Eve) are often unpredictable, further complicating the monitoring efforts. In contrast, machine learning-based methods systematically analyze and characterize attack patterns, enabling the detection of a broader spectrum of quantum attacks. Although existing studies \cite{Mao-2020, mao2020hidden, liao2022, Luo-2022, ding2023machine, kish2024mitigation} have demonstrated that machine learning models can strengthen the practical security of CV-QKD systems, a critical issue appears to have been overlooked: as CV-QKD systems advance from megahertz (MHz)- to terahertz (THz)-level rates \cite{he2020indoor, wang2022sub, ji2024gbps}, the corresponding surge in raw signal data significantly increases the training time of these models, rendering it nearly impractical---a bottleneck especially evident in high-rate, real-time applications.

In recent years, quantum machine learning (QML) \cite{biamonte2017quantum, shi2022parameterized, cerezo2021variational, cerezo2022challenges, tian2023recent, shi2024qsan, zhao2024qksan, wang2025interpretable} has emerged as a promising learning paradigm \cite{havlivcek2019supervised, banchi2021generalization}, leveraging quantum parallelism \cite{deutsch2020harnessing, wu2025quantum} and the computational power of high-dimensional Hilbert space \cite{Schuld2019quantum} to potentially accelerate classical machine learning tasks \cite{yin2025experimental}. For example, Ref.~\cite{lloyd2013quantum} proved that QML algorithms leverage efficient operations on high-dimensional vectors in tensor product spaces, which substantially reduce time complexity and lead to exponential speedups. Ref.~\cite{liu2021rigorous} formulated a well-defined classification problem and rigorously demonstrated that quantum kernel methods achieve an end-to-end exponential speedup. Collectively, these studies indicate that the QML algorithms may accelerate classical machine learning tasks and contribute to improved computational efficiency.

Inspired by the potential of the QML algorithms, we propose a QML-based attack detection framework (QML-ADF), which exploits QML models to identify quantum attacks exhibiting observable features. Specifically, we first design a collection of observable features that comprehensively characterize optical pulses subjected to quantum attacks. Then, we extract feature vectors from the optical pulses exchanged between the legitimate communication parties, Alice and Bob, and utilize these vectors as input for the QML models. Finally, the trained QML models are deployed to identify and predict the input data, with the corresponding prediction results determining whether the final secret key is generated.

The main contributions of this paper can be summarized as follows:
\begin{itemize}
	\item The QML-ADF is a purpose-built and systematically validated framework that addresses the unique security vulnerabilities of high-rate CV-QKD systems and, to the best of our knowledge, constitutes the first use of QML for attack detection in quantum communication.
	\item Twelve QML variants—six from quantum support vector machines (QSVM) and six from quantum neural networks (QNN)—are developed to underpin the QML-ADF. A comprehensive benchmarking and comparative analysis of twelve QML variants is further conducted, offering quantitative guidance for model selection.
	\item Three physically interpretable noise backends with varying noise intensities are constructed to evaluate the QML-ADF, and a comprehensive set of evaluation criteria confirms its exceptional robustness and practicality in detecting both known and previously undiscovered quantum attacks.
\end{itemize}

The structure of this paper is organized as follows. Section~\ref{relatedwork} provides an overview of the related work. Section~\ref{method} introduces the theoretical foundations of the QML-ADF. Section~\ref{framework} details the QML-ADF. In Sec.~\ref{experiments}, we evaluate the performance of the QML-ADF in detecting both known and previously undiscovered quantum attacks. Section~\ref{indepth} provides an in-depth analysis of the QML-ADF under various physically interpretable noise backends. Finally, Sec.~\ref{conclusion} concludes the paper with the main findings.

\section{Related work}\label{relatedwork}
Machine learning has been extensively applied to various complex tasks, including—but not limited to—high-dimensional classification, nonlinear regression, and real-time anomaly detection \cite{nie2018investigation, zhang2020empowering, huang2021secure, zhang2023survey, guo2023machine}. In recent years, a number of machine learning–based studies \cite{Mao-2020, mao2020hidden, liao2022, Luo-2022, ding2023machine, kish2024mitigation} have investigated defense strategies against various quantum attacks in CV-QKD systems. These strategies are principally built upon attack detection frameworks developed using classical machine learning methods, such as support vector machines \cite{ding2023machine}, decision trees \cite{kish2024mitigation}, and neural networks \cite{Mao-2020, Luo-2022}. Typically, such frameworks extract statistical or temporal features from raw signal data, which are then utilized to train discriminative models for identifying quantum attacks. For instance, Kish \textit{et al.} \cite{kish2024mitigation} proposed a lightweight and fast attack detection framework based on decision trees to identify diverse channel tampering attacks. However, this framework struggles to detect other types of quantum attacks. To overcome this limitation, Mao \textit{et al.} \cite{Mao-2020} introduced an artificial neural network (ANN)–based attack detection framework, designed to identify a broader range of quantum attacks. Furthermore, Liao \textit{et al.} \cite{liao2022} utilized the density-based spatial clustering of applications with noise (DBSCAN) algorithm \cite{shen2016real} for anomaly detection in CV-QKD systems. Inspired by this approach, Ding \textit{et al.} \cite{ding2023machine} developed a machine learning-based attack detection framework that combines DBSCAN with multiclass support vector machines (MCSVM) \cite{liu2015joint}, achieving excellent performance in identifying various quantum attacks.

Although these previously mentioned attack detection frameworks can identify a broader range of quantum attacks, they have a twofold impact on system performance. First, the number of optical pulses available for key extraction is reduced, as some must be sacrificed for shot-noise estimation, leading to a lower secret key rate. Second, the insertion loss introduced by optical switches or amplitude modulators further limits the maximum secure transmission distance. In contrast, the proposed QML-ADF introduces an auxiliary homodyne detector in the LO path to carry out real-time shot-noise analysis, thereby avoiding any impact on the secret key rate or the maximum secure transmission distance.

Furthermore, as transmission rates increase from the MHz to the THz scale \cite{he2020indoor, wang2022sub, ji2024gbps}, existing attack detection frameworks also encounter a dual dilemma: on one hand, the surge in data throughput significantly increases detection latency; on the other hand, the growing variety of quantum attacks—especially the emergence of previously unknown threats—leads to a marked decline in detection accuracy. Balancing high throughput, low latency, and high accuracy remains a fundamental challenge in the design of traditional attack detection frameworks.

QML has emerged as an alternative paradigm for addressing the limitations of classical machine learning methods, with the potential to achieve exponential computational advantages \cite{biamonte2017quantum, shi2022parameterized, cerezo2021variational, cerezo2022challenges, tian2023recent}. Recent developments in QML have largely concentrated on two foundational models: (i) QSVM leveraging quantum kernel estimation, adept at capturing high-dimensional and nonlinear patterns \cite{havlivcek2019supervised, Schuld2019quantum, yin2025experimental, liu2021rigorous, jerbi2023quantum, ding2024quantum, schuld2021supervised, wu2021application, thanasilp2024exponential}; and (ii) QNN formulated through variational quantum circuits, suitable for gradient-based optimization via backpropagation \cite{tian2023recent, shi2024qsan, zhao2024qksan, wang2025interpretable, wu2025quantum, mitarai2018quantum, farhi2018classification, abbas2021power, du2021learnability, larocca2023theory, schuld2020circuit, killoran2019continuous}. For instance, Havl\'{\i}\v{c}ek \textit{et al.} \cite{havlivcek2019supervised} demonstrated a QSVM architecture that utilizes quantum kernel estimation to perform proof-of-concept binary classification on superconducting quantum hardware. In Ref.~\cite{yin2025experimental}, Yin \textit{et al.} validated an optical QSVM architecture that leverages photonic quantum kernels to perform binary classification on a photonic processor. However, straightforward binary models cannot effectively handle multiclass classification. To this end, Ding \textit{et al.} \cite{ding2024quantum} proposed a novel QSVM framework, generalized for multiclass classification, which outperforms its classical counterpart in terms of performance. 

Furthermore, several pioneering studies have laid the foundation for QNN. In Ref.~\cite{mitarai2018quantum}, Mitarai \textit{et al.} introduced a QNN framework based on low-depth quantum circuits for function approximation, classification, and quantum many-body system simulation. In Ref.~\cite{schuld2020circuit}, Schuld \textit{et al.} proposed a QNN framework employing strongly entangled quantum circuits for supervised classification. Distinct from Ref.~\cite{mitarai2018quantum} and Ref.~\cite{schuld2020circuit}, Killoran \textit{et al.} \cite{killoran2019continuous} presented an optical continuous-variable QNN framework built from Gaussian and non-Gaussian gates, and demonstrated its effectiveness across classification, generative modeling, and hybrid learning tasks. Despite these promising advances, there exists no prior work exploring the application of QSVM or QNN to attack detection. To address the limitations of existing attack detection frameworks, we propose QML-ADF, which leverages either QSVM or QNN to identify both known and previously unknown quantum attacks without compromising the secret key rate or the maximum secure transmission distance.

\begin{figure*}[!t]
	\centering
	\includegraphics[width=0.6\linewidth]{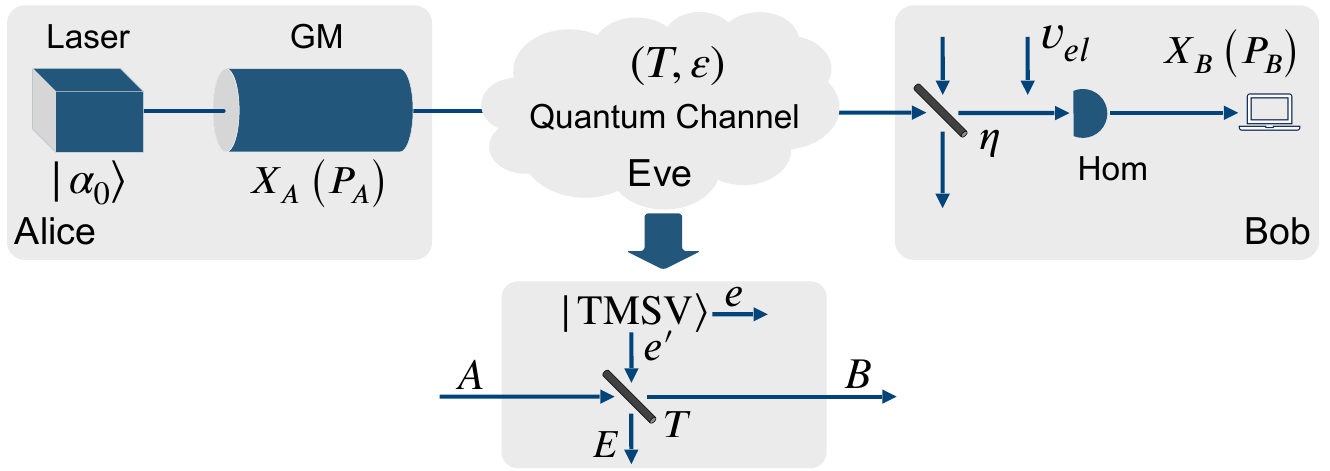}
	\caption{\textbf{CV-QKD scheme using Gaussian-modulated coherent states.} Alice’s mode $A$ is coupled to Eve’s mode $e^{\prime}$ via a beam splitter with transmittance $T$, resulting in output modes $B$, received by Bob, and $E$, retained by Eve. GM: Gaussian modulation; Hom: homodyne detection.}
	\label{figure01}
\end{figure*}

\section{Preliminaries}\label{method}
This section first introduces the CV-QKD mechanism, then reviews quantum feature maps.

\subsection{CV-QKD Mechanism}
Figure~\ref{figure01} illustrates a schematic of a CV-QKD scheme that transmits encrypted information using Gaussian-modulated coherent states. In this scheme, Alice encodes Gaussian random variables—drawn from a normal distribution with zero mean and variance $V_A$—onto the quadrature components $X_A$ and $P_A$ of the optical field via Gaussian modulation \cite{grosshans2003quantum}. The Gaussian-modulated coherent states are then transmitted to Bob via a quantum channel. At the receiving end of the quantum channel, Bob employs a homodyne detector—characterized by detection efficiency $\eta$ and electronic noise $v_{el}$—to measure the incoming quantum states, followed by basis reconciliation with Alice via a classical authenticated channel. Finally, through classical postprocessing, both parties share a secure secret key.

During transmission, the quantum channel may be subject to a collective Gaussian attack \cite{garcia2006unconditional} launched by Eve. In this scenario, Eve employs an entangling cloner to couple an eavesdropping module—consisting of a beam splitter with transmittance $T$ and a two-mode squeezed vacuum (TMSV) state $|\mathrm{TMSV}\rangle$ with variance $\mu$—into the quantum channel. The TMSV state possessed by Eve, comprising modes $e$ and $e^{\prime}$, has an associated covariance matrix \cite{weedbrook2012gaussian}
\begin{equation}
	V_{e e^{\prime}}=\left(\begin{array}{cc}
		\mu \mathbb{I} & \sqrt{\mu^2-1} \sigma_z \\
		\sqrt{\mu^2-1} \sigma_z & \mu \mathbb{I}
	\end{array}\right),
\end{equation}
where $\mathbb{I}=\operatorname{diag}(1,1)$ and $\sigma_z=\operatorname{diag}(1,-1)$. In addition, the excess noise $\varepsilon$ of the quantum channel is determined by $\mu$, satisfying the relation $\varepsilon={(\mu-1-T\mu+T)}/{T}$. At Bob's side, the homodyne detector yields a measurement outcome satisfying $y=\sqrt{\eta T} x + z$, where $z \sim \mathcal{N}\left(0, 1+v_{el}+\eta T \varepsilon\right)$ is the total noise term. Finally, the corresponding variance is given by $v_y = \eta T\left(V_{A} +\varepsilon\right)+ 1 + v_{el}$.

\subsection{Quantum Feature Maps}
To map classical data into quantum feature states, we explore the encoding schemes of two prominent quantum feature maps: angle encoding \cite{zhao2024qksan, ding2024quantum} and instantaneous quantum polynomial (IQP) encoding \cite{havlivcek2019supervised}. The circuit structures of angle encoding and IQP encoding are described in detail in Refs.~\cite{havlivcek2019supervised,ding2024quantum}. The angle encoding includes three circuit variants: AnRx, AnRy, and AnRz, corresponding to parameterized rotations along the X, Y, and Z axes, respectively. Therefore, we derive the following three quantum feature states:
\begin{equation}
	\!\!\!\!|\phi_{x}(\vec{x}_i)\rangle \!=\!\! \sum_{m_0\!=\!0}^1 \!\! \cdots\! \!\!\! \sum_{m_{n-1}\!=\!0}^1 \!\prod_{i=0}^{{n \!-\!1}}\!{\big(}\cos\frac{x_i}{2}{\big)}^{\!1\!-m_i}\!{\big(}\!-i \sin\frac{x_i}{2}{\big)}^{m_i}\! |m\rangle,
\end{equation}
\begin{equation}
	\!\!\!\!|\phi_{y}(\vec{x}_i)\rangle = \sum_{m_0\!=\!0}^1 \!\! \cdots \!\!\!\! \sum_{m_{n-1}\!=\!0}^1 \!\prod_{i=0}^{{n \!-\!1}}\!{\big(}\cos\frac{x_i}{2}{\big)}^{\!1\!-m_i}\!{\big(}\sin\frac{x_i}{2}{\big)}^{m_i}|m\rangle,
\end{equation}	
\begin{equation}
	|\phi_{z}(\vec{x}_i)\rangle = {\Big(}\frac{1}{\sqrt{2}}{\Big)}^{n} \sum_{m=0}^{2^{{n}}-1} \exp{{\Big(}i\sum_{i=0}^{{n-1}}x_i m_i {\Big)}} |m\rangle,
\end{equation}
where $m=m_{{n}-1} 2^0 + m_{{n}-2} 2^1 + \cdots + m_0 2^{{n}-1}$. Similarly, the IQP encoding comprises three circuit variants: IQPl, IQPc, and IQPf, which correspond to linear, circular, and full entanglement patterns, respectively. Accordingly, the following three quantum feature states can be established:
\begin{equation}
	\!\!\!\!|\phi_{l}(\vec{x}_i)\rangle \!=\! {\Big(}\frac{1}{\sqrt{2}}{\Big)}^{n} \! \sum_{m=0}^{2^{{n}}-1} \exp{{\Big(}i\sum_{i=0}^{{n-1}}x_i m_i{\Big)}}\!\! \prod_{(i, j) \in \bar{l}} \! \gamma^{\prime}_{(i,j)} |m\rangle, 
\end{equation}
\begin{equation}
	\!\!\!\!\!|\phi_{c}(\vec{x}_i)\rangle \!=\! {\Big(}\frac{1}{\sqrt{2}}{\Big)}^{n}\! \sum_{m=0}^{2^{{n}}-1} \exp{{\Big(}i\sum_{i=0}^{{n-1}}x_i m_i {\Big)}}\!\! \prod_{(i, j) \in \bar{c}} \! \gamma^{\prime}_{(i,j)}  |m\rangle, \\
\end{equation}	
\begin{equation}	
	\!\!\!\!|\phi_f(\vec{x}_i)\rangle \!=\! {\Big(}\frac{1}{\sqrt{2}}{\Big)}^{n}\! \sum_{m=0}^{2^{{n}}-1} \exp{{\Big(}i\sum_{i=0}^{{n-1}}x_i m_i {\Big)}}\!\! \prod_{(i, j) \in \bar{f}}  \! \gamma^{\prime}_{(i,j)} |m\rangle,
\end{equation}
where $\gamma^{\prime}_{(i,j) }   = \exp[{(-i {x_{i}x_{j}}/{2}) \left(-1\right)^{m_i \oplus m_j}}]$, $\bar{l}=\{(i, j) \mid 0 \leqslant i<j \leqslant n-1, j = i+1\}$, $\bar{c}=\{(i, j) \mid 0 \leqslant i \leqslant n-1, j = (i+1) \mod {n}\}$, and $\bar{f}=\{(i, j) \mid 0 \leqslant i<j \leqslant {n-1}\}$.

\begin{figure*}[!t]
	\centering 
	\includegraphics[width=0.80\linewidth]{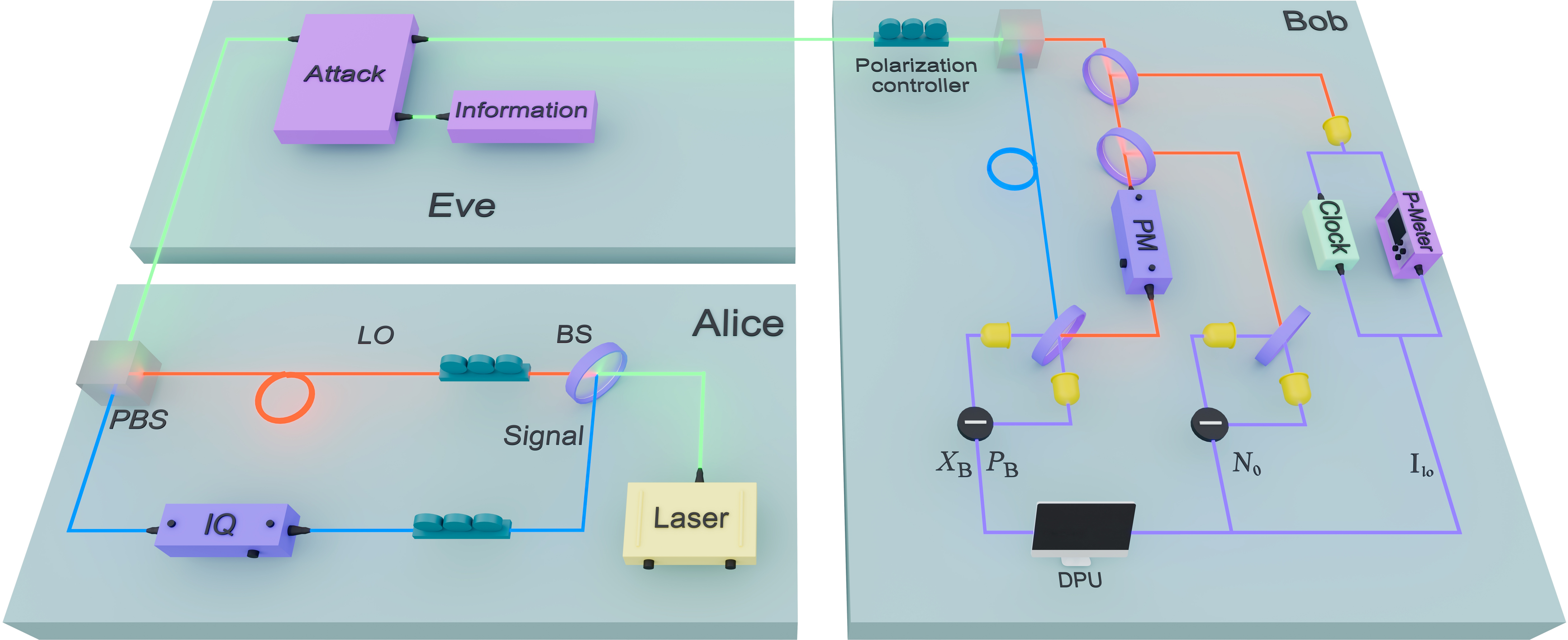}
	\caption{\textbf{Feature extraction and attack data collection.} During Alice's state preparation, a laser diode generates the initial pulses, which are split into a weak signal and a strong LO via a beam splitter (BS). In-phase and quadrature (IQ) modulation is applied on the signal path to prepare coherent states whose quadratures are modulated according to a Gaussian distribution. A delay line is employed on the LO path to synchronize the signal. Polarization beam splitters (PBS) are utilized for multiplexing and demultiplexing. During Bob’s measurement, a delay line on the signal path adjusts the signal timing, whereas the phase modulator (PM) on the LO path allows for random selection of the quadrature to be measured. A power meter (P-Meter) monitors the LO intensity, and a clock ensures precise synchronization. The first homodyne detector measures the received signal, whereas the second performs real-time shot-noise analysis. Polarization controllers optimize the polarization of both the signal and LO to maximize interference efficiency in the homodyne detection. Finally, the collected data ($X_B$, $P_B$, $N_0$, $I_{lo}$) is forwarded to the data processing unit (DPU) for attack detection and raw key distillation.}
	\label{figure06}
\end{figure*}

\section{quantum machine learning-based attack detection framework}\label{framework}
The proposed QML-ADF consists of three core modules: feature extraction, model architecture, and model inference, which are detailed below.
\begin{figure*}[!t]
	\centering
	\includegraphics[width=0.87\linewidth]{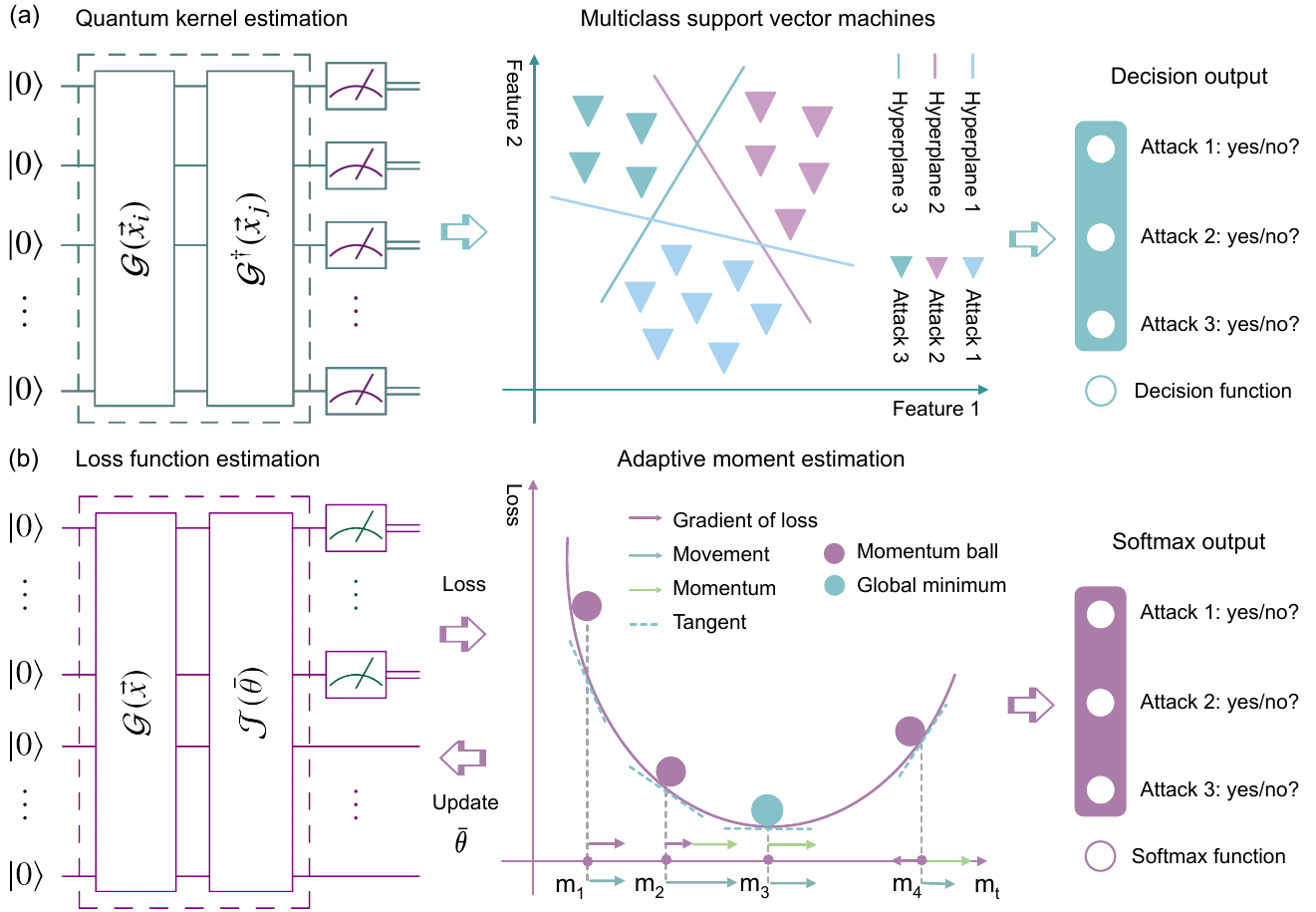}
	\caption{\textbf{QML models for noise-resistant and feature-aware attack detection.} (a) QSVM. The core of the model is quantum kernel estimation, which evaluates the fidelity between two quantum feature states. This is implemented by sequentially applying the unitary operation $\mathcal{G}(\vec{x}_i)$, its inverse $\mathcal{G}^\dagger(\vec{x}_j)$, and subsequently measuring all qubits at the output of the quantum circuit. (b) QNN. The model's performance is evaluated using a loss function defined over the trainable parameters $\bar{\theta}$. Optimization is performed via the classical Adam algorithm, which iteratively explores the parameter space defined by an ansatz $\mathcal{J}(\bar{\theta})$. At each iteration, Adam computes the loss with the current parameters and updates them according to adaptive estimates of first and second moments. This process is repeated until convergence to an optimal solution.}
	\label{figure07}
\end{figure*}

\subsection{Feature Extraction}
Figure~\ref{figure06} illustrates a schematic diagram of feature extraction from optical pulses. First, the sender, Alice, prepares a sequence of coherent states $|X_\textit{A} + i P_\textit{A}\rangle$ using Gaussian modulation \cite{grosshans2003quantum,grosshans2002continuous}. The quadrature components $X_\textit{A}$ and $P_\textit{A}$ are Gaussian distributed with zero mean and variance $V_AN_0$, where $N_0$ is the shot-noise variance. Then, the prepared coherent states, along with a strong LO, are transmitted to the receiver, Bob, through a quantum channel. Finally, Bob employs a homodyne detector to measure one quadrature component of the received coherent state and transmits the corresponding measurement basis (i.e., the selected quadrature component) to Alice over a classical authenticated channel. Following basis reconciliation, Alice and Bob collaboratively perform postprocessing to generate two correlated data sets, $\textbf{\textit{x}} =\{x^{\prime}_1, x^{\prime}_2, \dots, x^{\prime}_{t}\}$ and $\textbf{\textit{y}} = \{y^{\prime}_1, y^{\prime}_2, \dots, y^{\prime}_{t}\}$. As a result, the means and variances of $\textbf{\textit{x}}$ and $\textbf{\textit{y}}$ satisfy the following relations:
\begin{align}
	\begin{array}{ll}\bar{X}_\textit{A}=0, & V_x=V_{A} N_0 , \\ \\
		\bar{X}_\textit{B}=0, & V_y = \eta T\left(V_\textit{A} N_0+\xi\right)+N_0+V_{el},\end{array}
\end{align}
where $\xi = \varepsilon N_0$ and $V_{el} = v_{el} N_0$.

Under realistic conditions, quantum attacks on a CV-QKD system may disturb different observable features, such as the LO intensity ${I}_{lo}$, shot-noise variance $N_0$, as well as the mean $\bar{X}$, variance $V_y$, entropy $H_y$, and range $R_y$ of $\textbf{\textit{y}}$. Appendix illustrates the impact of quantum attacks on these observable features. It is worth noting that $V_y$, $H_y$, and $R_y$ are all clearly affected, whereas $\bar{X}$, ${I}_{lo}$, and $N_0$ experience varying levels of impact.~Therefore, systematically extracting and analyzing the observable features of optical pulses benefits the proposed QML-ADF in identifying and detecting complex quantum attacks. Suppose Bob receives a total of $g$ optical pulses, grouped into blocks of $g^{\prime}$ pulses each. The total number of blocks is $M = g/g^{\prime}$. For each block, we extract a feature vector $\vec{v} = (\bar{X}, V_y, I_{lo}, N_0, H_y, R_y)$ to comprehensively capture its distinctive characteristics. The resulting set of feature vectors $V = \{\vec{v}_1, \vec{v}_2, \dots, \vec{v}_M\}$ serves as the input to the QML model.

\subsection{Model Architecture}
Figure~\ref{figure07} depicts two QML models proposed for QML-ADF, designed to effectively identify and detect various quantum attacks. In the first model, we build upon quantum kernel estimation \cite{jerbi2023quantum, paine2023quantum, ding2024quantum} and employ MCSVM to construct decision hyperplanes. In the second model, QNN utilize a variational quantum circuit to learn data-dependent quantum feature representations. The circuit is followed by an adaptive measurement scheme to extract predictive information for attack detection. The adaptive measurement scheme dynamically adjusts the number of qubits measured based on the types and features of quantum attacks. For example, if three types of quantum attacks are detected, only three qubits need to be measured. As in classical neural networks, the QNN employ a classical adaptive moment estimation (Adam) \cite{kingma2014adam} method to accelerate convergence. Below is a detailed description of the two QML models.

\subsubsection{Quantum Support Vector Machines}
The QSVM model consists of two stages: (i) it computes a quantum kernel matrix using parameterized quantum circuits (PQC) on a quantum computer; and (ii) it solves quadratic programming problems using MCSVM (Fig.~\figref{figure07}{a}). In the initial stage, the input $\vec{x} \in \mathbb{R}^{n}$ is mapped to a quantum feature state $|\phi(\vec{x})\rangle$ through the application of quantum feature maps. In angle encoding, $\vec{x}$ is nonlinearly transformed into $|\phi_i(\vec{x})\rangle$ via the mapping 
\begin{equation}
	\phi_i: \vec{x} \mapsto |\phi_i(\vec{x})\rangle\langle\phi_i(\vec{x})| \in \mathbb{C}^{2^n \times 2^n}, \nonumber
\end{equation}
where $|\phi_i(\vec{x})\rangle$ denotes the quantum state prepared using rotational gates around the $i$-axis, with $i \in \{x, y, z\}$. Similarly, in IQP encoding, $\vec{x}$ is nonlinearly encoded into $|\phi_j(\vec{x})\rangle$ via the mapping
\begin{equation}
	\phi_j: \vec{x} \mapsto |\phi_j(\vec{x})\rangle\langle\phi_j(\vec{x})| \in \mathbb{C}^{2^n \times 2^n}, \nonumber
\end{equation}
where $|\phi_j(\vec{x})\rangle$ represents the state encoded using IQP circuits, with $j \in \{l, c, f\}$. Six QSVM variants are thus developed: three employing angle encoding (AnRx-QSVM, AnRy-QSVM, and AnRz-QSVM) and three utilizing IQP encoding (IQPl-QSVM, IQPc-QSVM, and IQPf-QSVM). The quantum kernel \cite{ding2024quantum} is then defined via the inner product:
\begin{equation}
	k(\vec{x}_i, \vec{x}_j) \operatorname{:}= |\langle \phi(\vec{x}_i) | \phi(\vec{x}_j) \rangle|^2,
\end{equation}
with $|\phi(\vec{x})\rangle \in \left\{ |\phi_a(\vec{x})\rangle \,\middle|\, a \in \{x, y, z, l, c, f\} \right\}$. Interestingly, we find that the quantum kernels constructed from quantum feature states $|\phi_{x}(\vec{x}_i)\rangle$, $|\phi_{y}(\vec{x}_i)\rangle$, and $|\phi_{z}(\vec{x}_i)\rangle$ are equivalent. A detailed proof can be found in Ref.~\cite{ding2024quantum}. Finally, the resulting quantum kernel matrix $\mathcal{Q}$ is formed by computing kernel values over all input data pairs, with entries $\mathcal{Q}_{ij} = k(\vec{x}_i, \vec{x}_j)$.
 
\begin{figure*}[!t]
	\centering
	\includegraphics[width=\linewidth]{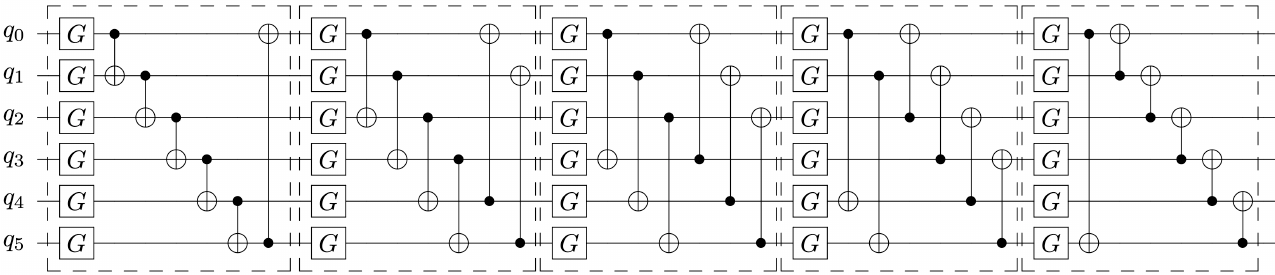}
	\caption{\textbf{Schematic of variational quantum circuits.}}
	\label{figure04}
\end{figure*}

In the second stage, the quantum kernel matrix $\mathcal{Q}$ is integrated into MCSVM to identify various quantum attacks. Typically, the MCSVM employ a one-against-the-rest strategy \cite{hsu2002comparison} to address quadratic programming problems. In this work, a $k$-class attack detection task corresponds to solving $k$ quadratic programming problems. We begin with a labeled dataset $\{(\vec{x}_i, y_i)\}_{i=1}^d$, where each $\vec{x}_i \in \mathbb{R}^n$ is a feature vector and $y_i \in \{1, 2, \dots, k\}$ indicates the class to which $\vec{x}_i$ belongs. The dual optimization problem for the $s$th quadratic programming problem is expressed as follows:
\begin{equation}\label{e2}
	\begin{aligned}
		&\min \enspace \frac{1}{2} \sum_{i, j=1}^{d}  \alpha_i^{s} \alpha_j^{s} y_i^{s} y_j^{s} {\Big(}|\langle \phi(\vec{x}_i)|\phi(\vec{x}_j) \rangle|^2 {\Big)} -\sum_{i=1}^{d} \alpha_i^{s}, \\ 
		&\textrm{such that}\begin{cases} \
			\sum_{i=1}^{d} \alpha_i^{s} y_i^{s} =0,	\\
			\ 0 \leqslant \alpha_i^{s} \leqslant C, \enspace i=1, 2, \dots, d,  \\
		\end{cases}
	\end{aligned}
\end{equation}
where $\alpha_i^{s}$ and $\alpha_j^{s}$ denote the Lagrange multipliers, and $C$ represents the penalty parameter. We assign $ y_i^{s} = +1 $ if the original label $y_i$ belongs to class $s$; otherwise, we set $y_i^{s} = -1$. The solution to the dual optimization problem yields the optimal Lagrange multiplier vector $\vec{\alpha}^{s} = {\big(}\hat{\alpha}_1^{s}, \hat{\alpha}_2^{s}, \dots, \hat{\alpha}_d^{s}{\big)}$, along with the decision function for class $s$:
\begin{equation}
	\tilde{f}^s{\left(\vec{p}\right)} = \sum_{i=1}^{d} {\hat{\alpha}}_i^{s}y_i^{s} 
	{\Big(}|\langle \phi(\vec{x}_i)|\phi(\vec{p}) \rangle|^2 {\Big)} + b_s,
\end{equation}
where $\vec{p}$ denotes the input sample and $b_s$ is the bias term. Consequently, the QSVM model defines $k$ decision functions, each producing a prediction score corresponding to one of the $k$ target classes. During inference, the predicted label corresponds to the class with the highest score. Formally, the final decision function is defined as
\begin{equation}\label{eq:fin_deci}
	\tilde{f}(\vec{p}) \operatorname{:}= \operatorname*{arg\,max}_{s \in \{1, \dotsc, k\}} \tilde{f}^s(\vec{p}),
\end{equation}
where $\tilde{f}^s(\vec{p})$ denotes the score for class $s$ given input $\vec{p}$.

\subsubsection{Quantum Neural Networks}
The QNN model is structured into five sequential steps. First, the input $\vec{x}\in \mathbb{R}^{n}$ is mapped to a quantum feature state in the same manner as in QSVM, as illustrated in Fig.~\figref{figure07}{b}. Here, we define six QNN variants: three based on angle encoding (AnRx-QNN, AnRy-QNN, and AnRz-QNN), and three based on IQP encoding (IQPl-QNN, IQPc-QNN, and IQPf-QNN).

Second, a variational quantum circuit $\mathcal{J}(\bar{\theta})$ is employed to the quantum feature state $|\phi(\vec{x})\rangle$ (Fig.~\ref{figure04}). The circuit is composed of $L_d$ trainable variational layers and is parameterized by the union $\bar{\theta}$ of its constituent parameter subvectors, which are optimized using Adam method. The variational quantum circuit $\mathcal{J}(\bar{\theta})$ is then given by
\begin{equation}
	\mathcal{J}(\bar{\theta})=\mathcal{J}_{L_d}\left(\bar{\theta}_{L_d}\right) \mathcal{J}_{{L_d}-1}\left(\bar{\theta}_{{L_d}-1}\right) \ldots \mathcal{J}_1\left(\bar{\theta}_1\right).
\end{equation}
For each unitary operation $\mathcal{J}_l(\bar{\theta}_l)$, there are $n$ trainable rotation gates and CNOT gates. The trainable rotation gate is denoted by
\begin{equation}
	\!\!\!\! G =\begin{pmatrix}
		\exp({\frac{-i(\phi+\lambda)}{2}}) \cos \left(\frac{\omega}{2}\right) & -\exp(\frac{i(\phi-\lambda)}{2}) \sin \left(\frac{\omega}{2}\right) \\
		\exp(\frac{-i(\phi-\lambda)}{2}) \sin \left(\frac{\omega}{2}\right) & \exp(\frac{i(\phi+\lambda)}{2}) \cos \left(\frac{\omega}{2}\right)
		\end{pmatrix}.
\end{equation}
In each CNOT gate within the unitary operation $\mathcal{J}_l(\bar{\theta}_l)$, the $q$th qubit is the control, and the target qubit is indexed by $q_c = (q+l) \mod n$, where $q \in [0, n-1]$ and $l \in [1, n-1]$.

Third, for a $k$-class attack detection task, an adaptive measurement scheme in the Z-basis is conducted on the quantum state $\mathcal{J}(\bar{\theta})|\phi(\vec{x})\rangle$. In other words, the adaptive measurement targets the Pauli-Z operator expectation values of the $k$ qubits. The Pauli-Z operator expectation value for each qubit is given by
\begin{equation}
	\langle {Z}_{q} \rangle = \langle \phi(\vec{x})| \mathcal{J}^{\dagger}(\bar{\theta}){Z}_{q}\mathcal{J}(\bar{\theta}) |\phi(\vec{x})\rangle =\delta_{q}^{(0)} - \delta_{q}^{(1)},
\end{equation}
where $\delta_{q}^{(0)}$ and $\delta_{q}^{(1)}$ denote the probabilities that the measurement outcome of the $q$th qubit is $|0\rangle$ and $|1\rangle$, respectively. The measurement then yields $\vec{Z} = {\big(} \langle  {Z}_{0} \rangle, \langle {Z}_{1} \rangle, \dots, \langle {Z}_{k-1} \rangle{\big)}$.

Fourth, the unnormalized measurement outcome $\vec{Z}$ is transformed into a normalized probability distribution $\vec{\delta}={\big(}\delta_0, \delta_1, \dots, \delta_{k-1}{\big)}$ by applying a softmax function
\begin{equation}
	\delta_q = \frac{\exp {{\Big(}\langle {Z}_q \rangle-\max_{q} \langle{Z}_q\rangle {\Big)}}}{\sum_{m=0}^{k-1} \exp{{\Big(}\langle {Z}_m \rangle\!-\!\max_{q} \langle{Z}_q\rangle  {\Big)}}}.
\end{equation}
The probability distribution $\vec{\delta}$ is suitable for calculating a cross-entropy loss function, which is expressed as
\begin{equation}
	\mathcal{L}_\mathrm{ce}=\max_{q} \langle{Z}_q\rangle \!-\! \langle {Z}_{s} \rangle \!+\! \log \sum_{m=0}^{k-1} \exp {\Big(}\langle{Z}_{m}\rangle \!-\! \max_{q}\langle{Z}_q\rangle{\Big)},
\end{equation}
where $\langle {Z}_{s} \rangle$ is the logit (i.e., the unnormalized score) associated with the ground-truth class.

Fifth, to enable backpropagation, we employ the chain rule in conjunction with a parameter-shift rule \cite{schuld2019evaluating} to efficiently estimate the gradient $\nabla_{\bar{\theta}} \mathcal{L}_{\mathrm{ce}}$. Therefore, we evaluate the gradient of $\mathcal{L}_\mathrm{ce}$ with respect to $\langle {Z}_{q} \rangle$, as well as the gradient of $\langle {Z}_{q} \rangle$ with respect to $\bar{\theta}$, denoted as ${\partial \mathcal{L}_\mathrm{ce}}/{\partial \langle {Z}_{q} \rangle}$ and $\nabla_{\bar{\theta}} \langle {Z}_{q}(\bar{\theta}) \rangle$, respectively. Specifically, we first calculate the gradient ${\partial \mathcal{L}_\mathrm{ce}}/{\partial \langle {Z}_{q} \rangle}$. This calculation considers two distinct cases. (i) $q = s$. The gradient of $\mathcal{L}_\mathrm{ce}$ with respect to $\langle {Z}_{s} \rangle$ is expressed as ${\partial \mathcal{L}_\mathrm{ce}}/{\partial \langle {Z}_{s} \rangle}  = \delta_s -1$. (ii) $q \neq s$. The gradient of $\mathcal{L}_\mathrm{ce}$ with respect to $\langle {Z}_{q} \rangle$ becomes ${\partial \mathcal{L}_\mathrm{ce}}/{\partial \langle {Z}_{q} \rangle} = \delta_q$. Suppose $y$ is the true class label, denoted by a one-hot encoded vector $\vec{y} = (y_0, y_1, \ldots, y_{k-1})$, where $y_s=1$ indicates the true class and all other entries are $0$. Therefore, we have ${\partial \mathcal{L}_\mathrm{ce}}/{\partial \langle {Z}_{q} \rangle} = \delta_q-y_q$. The gradient of $\mathcal{L}_\mathrm{ce}$ with respect to $\bar{\theta}$ is then computed via the chain rule, as follows:
\begin{equation}
	\nabla_{\bar{\theta}} \mathcal{L}_\mathrm{ce} = \sum_{q=0}^{k-1} \left(\delta_q-y_q \right) \cdot \nabla_{\bar{\theta}} \langle {Z}_{q}(\bar{\theta}) \rangle.
\end{equation}
While the gradient ${\partial \mathcal{L}_\mathrm{ce}}/{\partial \langle Z_{q} \rangle}$ is relatively straightforward to compute in classical settings, evaluating $\nabla_{\bar{\theta}} \langle Z_{q}(\bar{\theta}) \rangle$---which depends on the partial derivatives of PQC—poses greater challenges. Here, the gradient $\nabla_{\bar{\theta}} \langle {Z}_{q}(\bar{\theta}) \rangle$ is typically evaluated using the parameter-shift rule \cite{schuld2019evaluating}. According to this rule, the derivative with respect to each circuit parameter can be obtained by computing expectation values at shifted parameter values. To minimize the cross-entropy loss, the parameters are iteratively updated following the Adam method. During training, this iterative optimization progressively adjusts the softmax output to better match the target class label.

\begin{figure}[!t]
	\centering
	\includegraphics[width=0.82\linewidth]{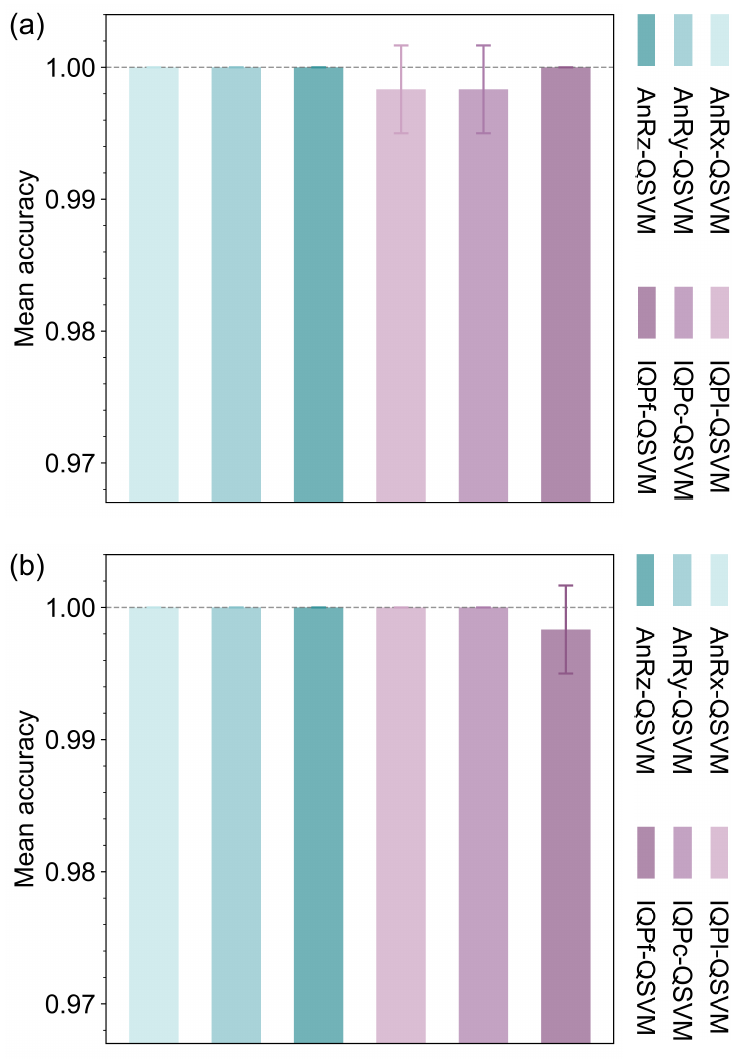}
	\caption{\textbf{Performance comparison across different QSVM variants.} (a) Mean accuracy comparison of six QSVM variants on the known quantum attack dataset. (b) Mean accuracy comparison of the same six variants on the unknown quantum attack dataset. Each dataset is partitioned into five folds for five-fold cross-validation, where in each iteration, four folds are used for training and one for testing. Accuracy is recorded for each iteration, and the final mean accuracy is computed by averaging the results across all five folds.}
	\label{figure08}
\end{figure}

\subsection{Model Inference}
As outlined above, the QML models employed in this work are QSVM and QNN. In noise-resistant feature-aware attack detection, model inference with either QSVM or QNN consists of the following steps. First, the input $V$ is standardized to have zero mean and unit variance. The resulting data $V^{\prime}$ are then fed into the trained QML model. In QSVM, the model evaluates the decision function values for each class and assigns the class with the highest value as the predicted label. In contrast, QNN employ PQC to generate an output vector based on measurement outcomes. This vector is then passed through a softmax function to produce a probability distribution over all classes, and the class with the highest probability is assigned as the predicted label. The final label determines whether the CV-QKD system is under quantum attacks. If no anomaly is detected, Alice and Bob proceed to generate and share a string of raw keys. To derive the final secret key, classical postprocessing is applied to the raw key string, including parameter estimation, reverse reconciliation, and privacy amplification.

\begin{figure}[b!]
	\centering
	\includegraphics[width=0.75\linewidth]{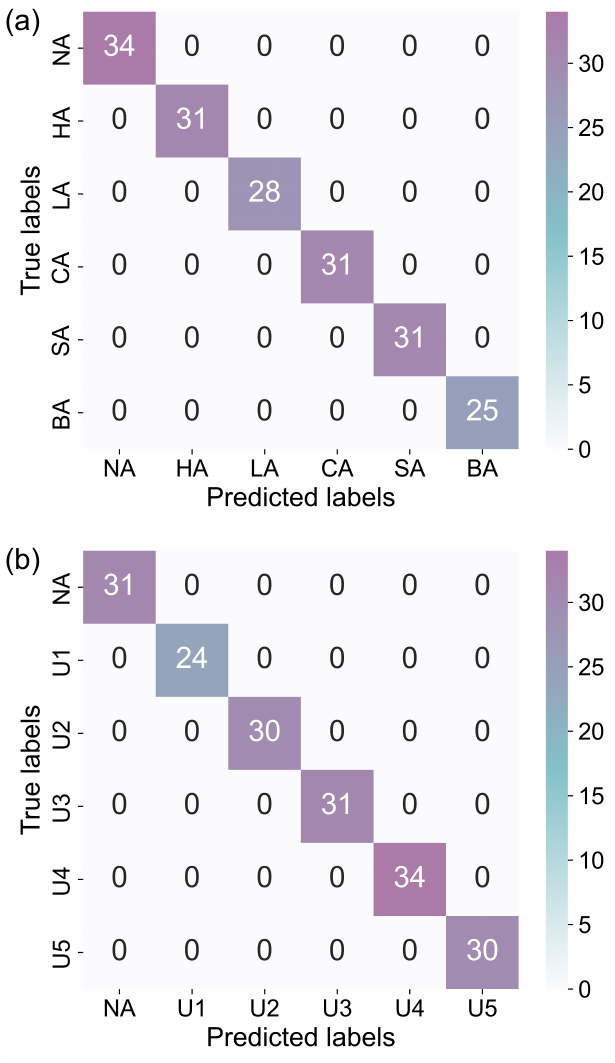}
	\caption{\textbf{Performance evaluation of AnR-Type QSVM.} (a–b) Confusion matrices for AnR-Type QSVM evaluated on the known (a) and unknown (b) quantum attack datasets. Both datasets are divided into training and testing sets following a $70$:$30$ ratio (see Appendix). NA: no attack; HA: homodyne-detector-blinding attack; LA: LO-intensity attack; CA: calibration attack; SA: saturation attack; BA: blended attack; U1--U5, unknown attack scenarios $1$ to $5$.}
	\label{figure09}
\end{figure}

\begin{figure}[!t]
	\centering
	\includegraphics[width=\linewidth]{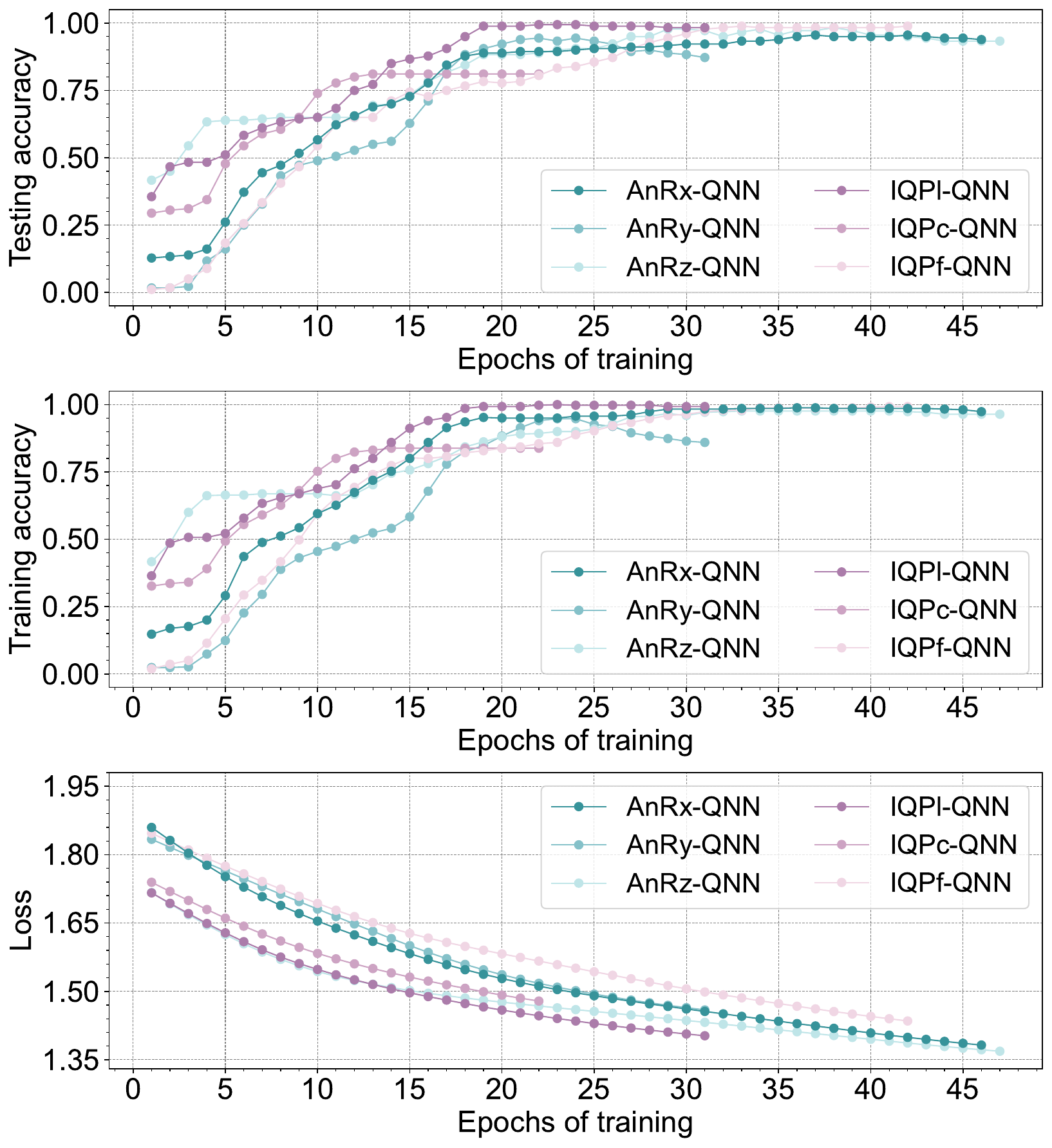}
	\caption{\textbf{Accuracy and loss functions of six QNN variants on the known quantum attack dataset.} Accuracy increases and plateaus as training progresses, while loss decreases and stabilizes.}
	\label{figure10}
\end{figure}
\begin{figure}[!t]
	\centering
	\includegraphics[width=\linewidth]{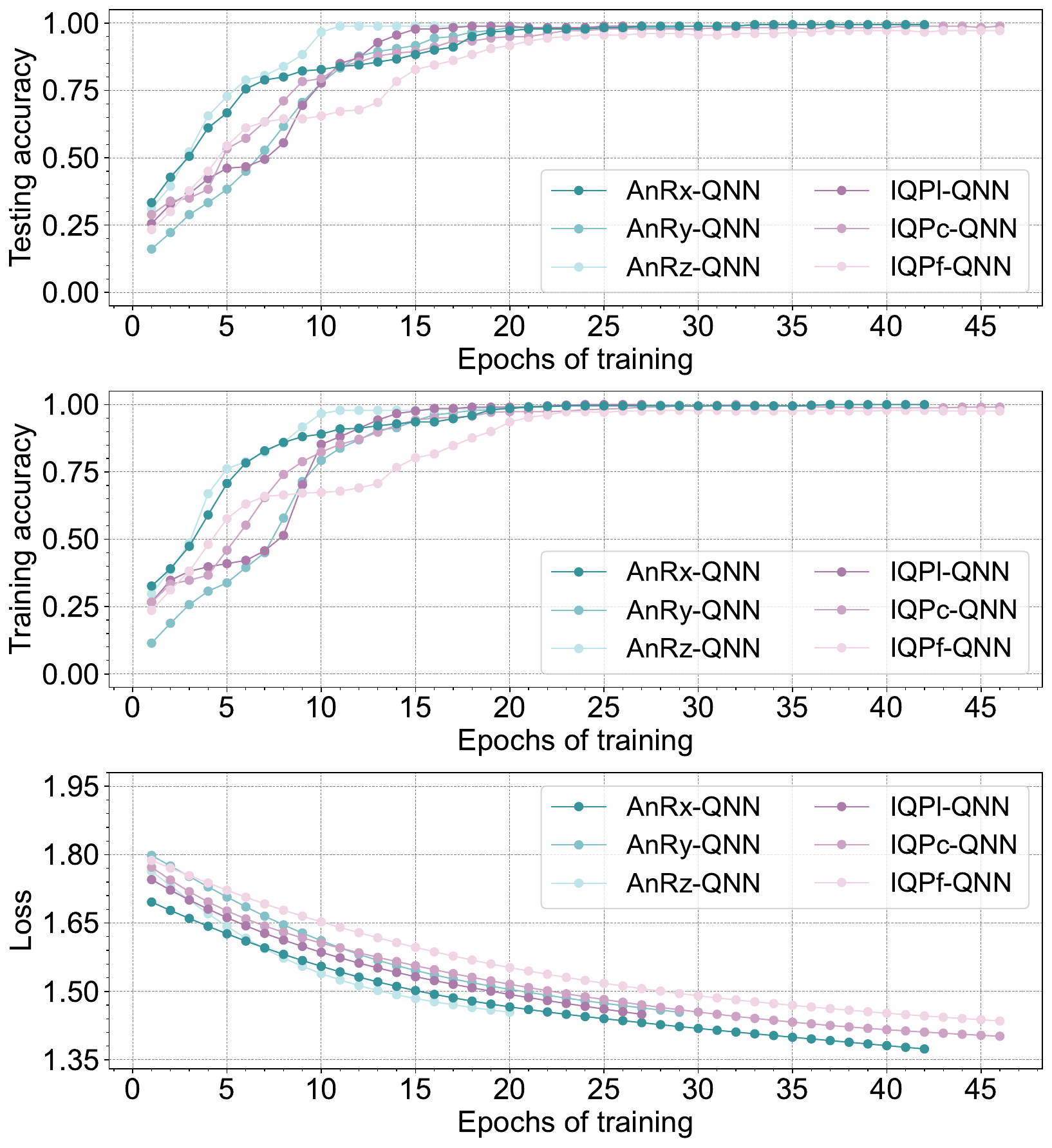}
	\caption{\textbf{Accuracy and loss functions of six QNN variants on the unknown quantum attack dataset.} Accuracy rises and levels off with training, while loss decreases and stabilizes.}
	\label{figure11}
\end{figure}

\section{Performance Benchmarking}\label{experiments}
This section benchmarks various QSVM and QNN variants under known and unknown quantum attacks. In this work, all training and testing tasks are performed on a Windows Server $2022$ system running $\textsc{python}$ (version $3.12.2$). The server is equipped with an $\text{Intel}64$ processor ($\text{Intel}64$ Family $6$ Model $85$ Stepping $7$, GenuineIntel), featuring $40$ physical cores, $80$ logical threads, $1021.64$ $\mathrm{GB}$ of RAM, and a $64$-bit architecture. In addition, quantum simulations of pure and mixed quantum systems are conducted using the $\textsc{default.qubit}$ and $\textsc{default.mixed}$ simulators, respectively \cite{bergholm2018pennylane}. In the following, QSVM and QNN variants are benchmarked to evaluate their resilience against both known and unknown quantum attacks.
\begin{figure}[!t]
	\centering
	\includegraphics[width=0.75\linewidth]{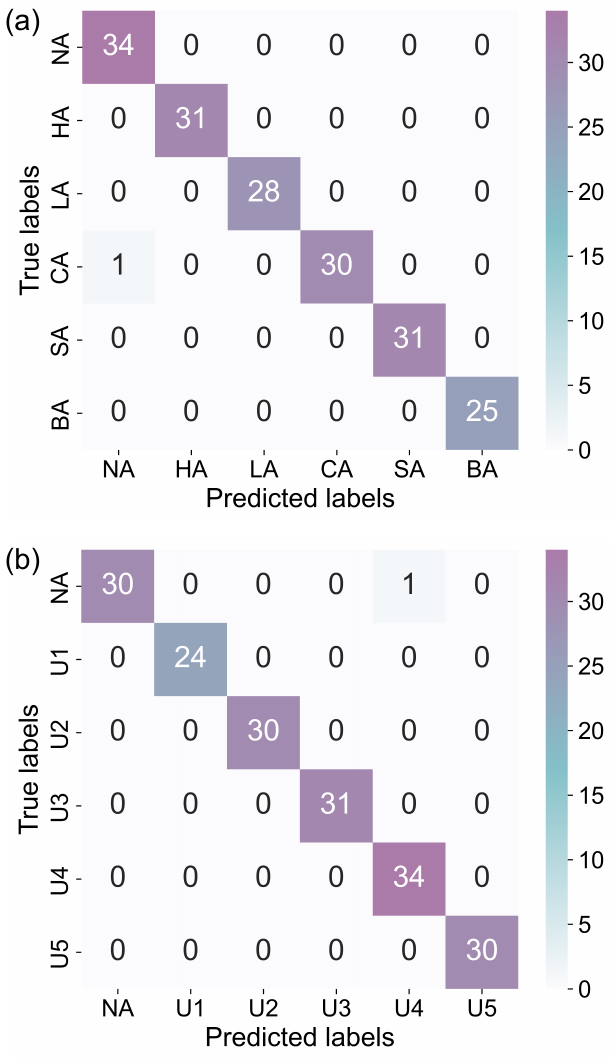}
	\caption{\textbf{Performance evaluation of IQPl-QNN and AnRx-QNN.} (a–b) Confusion matrices of IQPl-QNN on the known quantum attack dataset (a) and of AnRx-QNN on the unknown dataset (b). Both datasets are split into training and testing sets using a $70$:$30$ ratio (see Appendix).}
	\label{figure12}
\end{figure}

\subsection{Benchmarking QSVM Variants}
Figure~\ref{figure08} compares the performance of various QSVM variants. Among them, AnRx-, AnRy-, and AnRz-QSVM demonstrate a marked advantage in detecting quantum attacks over the other QSVM variants. In addition, these three variants yield nearly identical performance in securing CV-QKD systems, owing to the equivalence of their underlying quantum kernels. Hereafter, we collectively refer to them as AnR-type QSVM. To further evaluate the performance of AnR-type QSVM, confusion matrices are employed to assess their attack detection across all classes. As illustrated in Fig.~\ref{figure09}, the AnR-type QSVM achieve perfect detection accuracy on both known and unknown quantum attack detection tasks.

\subsection{Benchmarking QNN Variants}
Figures~\ref{figure10} and \ref{figure11} illustrate the performance of various QNN variants on the known and unknown quantum attack datasets, respectively. To prevent overfitting, a fixed early stopping criterion is applied to all QNN variants, under which improvements on the testing set cease after $10$ epochs. Moreover, Adam is used with an initial learning rate of $1 \times 10^{-3}$, and a learning rate scheduler is used to progressively refine it during training. The scheduler is configured with a patience of $5$ epochs, a decay factor of $0.5$, and a minimum learning rate threshold of $1 \times 10^{-6}$. With these experimental parameters established, we proceed to evaluate the performance of various QNN variants.

As shown in Fig.~\ref{figure10}, IQPl-QNN markedly outperform the other QNN variants in the known quantum attack detection task. In contrast, Fig.~\ref{figure11} shows that AnRx-QNN exhibit superior performance over the other QNN variants in the unknown quantum attack detection task. To further evaluate the performance of IQPl-QNN on the known quantum attack dataset and AnRx-QNN on the unknown dataset, confusion matrices are used to assess their attack detection across all classes. Figure~\ref{figure12} illustrates the confusion matrices of IQPl-QNN and AnRx-QNN. As the results indicate, IQPl-QNN demonstrate near-perfect detection accuracy on the known quantum attack detection task, with only one misdetection. Similarly, AnRx-QNN make just one error on the unknown task, yielding an accuracy close to $100\%$. Therefore, IQPl-QNN and AnRx-QNN each exhibit excellent attack detection performance in their respective tasks.
\begin{table*}[!t]
	\centering
	\caption{\label{table02}\textbf{Design of physically interpretable noise backends based on noise metrics of representative superconducting quantum hardware.}}
	\begin{tabular*}{\linewidth}{@{\extracolsep{\fill}}l@{}c@{}c@{}c@{}c@{}c@{}c@{}c@{}c@{}}
		\toprule
		Provider &Backend & $T_1 (\SI{}{\micro\second})$ & $T_2 (\SI{}{\micro\second})$ & $\epsilon_1$ & $\epsilon_\mathrm{m}$ ($\approx \delta_\mathrm{BFC}$) & $\delta_\mathrm{ADC}$ & $\delta_\mathrm{PDC}$ & $\delta_\mathrm{DPC}$ \\
		\midrule
		IBM &$\text{ibm\_aachen}$ \cite{ibm_backend} & $2.17 \times 10^{2}$ & $1.84 \times 10^{2}$ & $2.19 \times 10^{-4}$ &$8.55 \times 10^{-3}$ & $4.61 \times 10^{-4}$ & $6.26 \times 10^{-4}$ & $3.29 \times 10^{-4}$ \\
		IBM &$\text{ibm\_marrakesh}$ \cite{ibm_backend}  & $2.04 \times 10^{2}$ & $9.72 \times 10^{1}$ & $\SI{3.48e-4}{}$ &$\SI{8.55e-3}{}$ & $4.90 \times 10^{-4}$ & $1.57 \times 10^{-3}$ & $5.22 \times 10^{-4}$ \\
		IBM &$\text{ibm\_torino}$ \cite{ibm_backend}  & $1.72 \times 10^{2}$ & $1.36 \times 10^{2}$ & $\SI{3.09e-4}{}$ &$\SI{2.25e-2}{}$ &$5.81 \times 10^{-4}$  & $8.89 \times 10^{-4}$  & $4.64 \times 10^{-4}$\\
		$\text{Google}$ &$\text{Willow}$ \cite{acharya2025quantum}  & $7.30 \times 10^{1}$ &  $8.00 \times 10^{1}$ &$\SI{6.20e-4}{}$ &$\SI{8.00e-3}{}$  & $1.37 \times 10^{-3}$  & $1.13 \times 10^{-3}$  & $9.30 \times 10^{-4}$ \\
		IQM & \text{Garnet} \cite{iqm_garnet}   &$\SI{4.01e+1}{}$ & $9.03 \times 10^{0}$ & $8.00 \times 10^{-4}$ &$\SI{3.20e-2}{}$ & $2.49 \times 10^{-3}$   &  $1.95 \times 10^{-2}$   & $1.20 \times 10^{-3}$ \\
		Rigetti & \text{Ankaa-3} \cite{rigetti_ankaa3}  & $\SI{3.30e+1}{}$ & $\SI{2.00e+1}{}$ &$8.00 \times 10^{-4}$ &$\SI{3.50e-2}{}$  &$3.03 \times 10^{-3}$   & $6.95 \times 10^{-3}$ & $1.20 \times 10^{-3}$   \\
		Ours & \text{low\_noise\_realistic\_model}  & $\SI{1.00e+1}{}$ & $\SI{1.00e+1}{}$ & $6.00 \times 10^{-3}$ &${6.75 \times 10^{-2}}$
		&  $9.95 \times 10^{-3}$ &$9.95 \times 10^{-3}$    & $9.00 \times 10^{-3}$\\
		Ours & \text{mid\_noise\_stress\_test\_model}  & $1.00 \times 10^{0}$ & $1.00 \times 10^{0}$ & $7.60 \times 10^{-2}$ & $\SI{9.25e-2}{}$&  $9.52 \times 10^{-2}$ &$9.52 \times 10^{-2}$   & $1.14 \times 10^{-1}$ \\
		Ours & \text{high\_noise\_adversarial\_model}  & $1.00 \times 10^{0}$ & $1.00 \times 10^{0}$ & $\SI{1.50e-1}{}$ &$\SI{1.50e-1}{}$&  $9.52 \times 10^{-2}$  &$9.52 \times 10^{-2}$      & $2.25 \times 10^{-1}$  \\
		\bottomrule
	\end{tabular*}
\end{table*}

Based on the above analysis, AnR-type QSVM exhibit state-of-the-art performance in identifying both known and previously unseen quantum attacks. Therefore, the proposed QML-ADF demonstrates excellent detection performance across various quantum attacks, thereby reinforcing the practical security of CV-QKD systems.

\section{Performance Benchmarking Regarding Robustness}\label{indepth}
So far, we have investigated the detection of quantum attacks under ideal, noise-free conditions, with a particular focus on the behavior of PQC in such settings. However, when implementing PQC on real quantum hardware, it is crucial to account for the effects of hardware-induced noise. In this work, we focus on four representative types of hardware noise channel: (i) bit-flip channel (BFC), (ii) amplitude-damping channel (ADC), (iii) phase-damping channel (PDC), and (iv) depolarizing channel (DPC). These noise channels stem from distinct physical error mechanisms and can be rigorously represented by a set of Kraus operators $\{M_s\}$ that satisfy the conditions of complete positivity and trace preservation (CPTP):
\begin{equation}
	\mathcal{M}(\rho)=\sum_s M_s \rho M_s^{\dagger}, \ \text{with} \ \sum_s M_s^{\dagger} M_s=\mathbb{I}.
\end{equation}

To accurately simulate hardware-induced noise, the noise metrics provided by real quantum hardware—such as relaxation time $T_1$, dephasing time $T_2$, single-qubit gate error rate $\epsilon_1$, and readout error $\epsilon_\mathrm{m}$—should be mapped to their corresponding noise channel parameters. Below, we investigate the relationship between each of the four noise channel parameters and their corresponding hardware noise metrics. Furthermore, we construct physically interpretable noise backends tailored for state-of-the-art AnR-type QSVM.
\begin{figure}[!b]
	\centering
	\includegraphics[width=\linewidth]{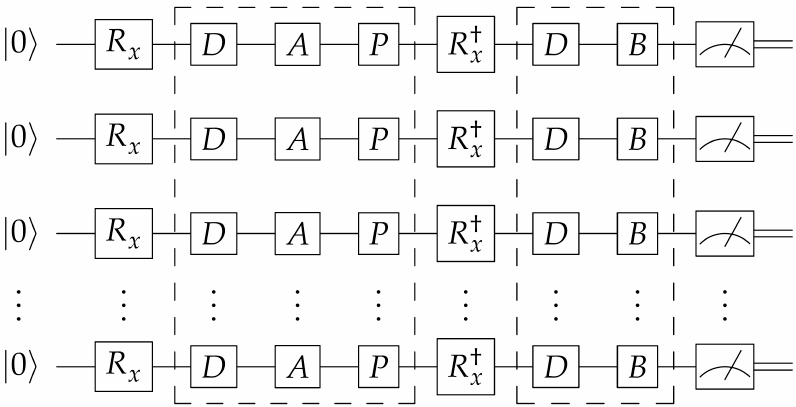}
	\caption{\textbf{AnRx-QSVM under physically interpretable noise backends.}}
	\label{figure13}
\end{figure}
\begin{figure}[!b]
	\centering
	\includegraphics[width=\linewidth]{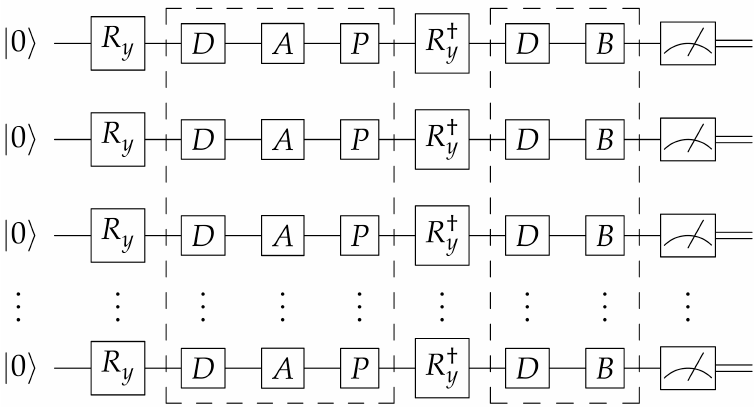}
	\caption{\textbf{AnRy-QSVM under physically interpretable noise backends.}}
	\label{figure14}
\end{figure}
\begin{figure*}[!t]
	\centering
	\includegraphics[width=\linewidth]{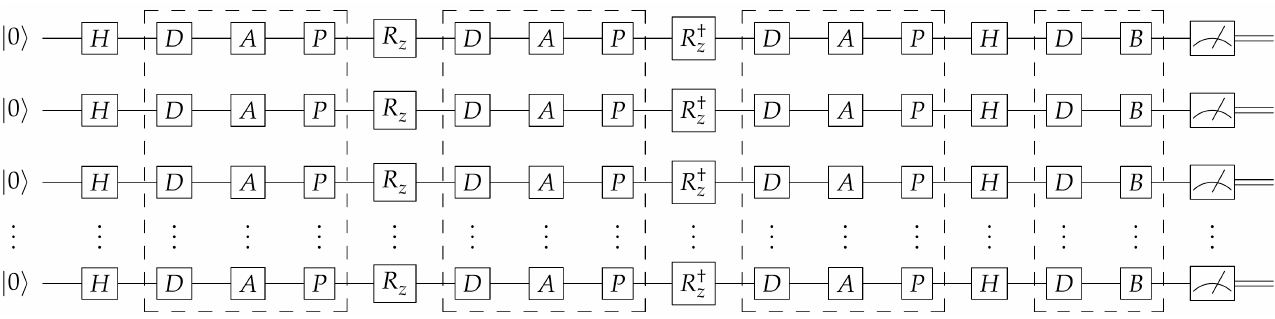}
	\caption{\textbf{AnRz-QSVM under physically interpretable noise backends.}}
	\label{figure15}
\end{figure*}

\subsection{Modeling Physically Interpretable Noise Backends}
\subsubsection{Bit-Flip Channel}
The BFC describes a noise mechanism where a qubit state is flipped between $\lvert 0 \rangle$ and $\lvert 1 \rangle$ with probability $\delta_\mathrm{BFC}$, and left unchanged with probability $1 - \delta_\mathrm{BFC}$. For a single-qubit system, the corresponding Kraus operator representation has the form:
\begin{equation}
	\begin{aligned}
		M_0^{(\mathrm{BFC})} &= \sqrt{1 - \delta_\mathrm{BFC}}\, \mathbb{I}, \\  M_1^{(\mathrm{BFC})} &= \sqrt{\delta_\mathrm{BFC}}\, \sigma_x,
	\end{aligned}
\end{equation}
where $\sigma_x=|0\rangle\langle 1|+|1\rangle\langle 0|$.
In this work, we model the readout error $\epsilon_\mathrm{m}$ using a bit-flip channel, i.e., $\delta_\mathrm{BFC} \approx \epsilon_\mathrm{m}$, where $\epsilon_\mathrm{m}$ denotes the average probability of measurement error observed in real quantum hardware.

\subsubsection{Amplitude-Damping and Phase-Damping Channels}
The ADC models irreversible energy loss in a quantum system, typically arising from spontaneous emission or relaxation to the ground state $\lvert 0 \rangle$. For a single-qubit system, the corresponding Kraus operator representation of this channel is given by:
\begin{equation}\label{adc}
	\begin{aligned}
		M_0^{(\mathrm{ADC})} &= |0\rangle\langle 0| + \sqrt{1-\delta_\mathrm{ADC}}|1\rangle\langle 1|,\\
		M_1^{(\mathrm{ADC})} &= \sqrt{\delta_\mathrm{ADC}} |0\rangle\langle 1|,
	\end{aligned}
\end{equation}
where $\delta_\mathrm{ADC} \in [0,1]$ represents the probability that the excited state $\lvert 1 \rangle$ decays to the ground state $\lvert 0 \rangle$ during the noise process. When $\delta_\mathrm{ADC} = 0$, no noise occurs; when $\delta_\mathrm{ADC} = 1$, $\lvert 1 \rangle$ always decays to $\lvert 0 \rangle$.

The PDC models the loss of quantum coherence resulting from interactions between the quantum system and its environment, without any associated energy dissipation. For a single-qubit system, the corresponding Kraus operator representation of this channel is expressed as:
\begin{equation}
	\begin{aligned}
		M_0^{(\mathrm{PDC})} &= |0\rangle\langle 0| + \sqrt{1-\delta_\mathrm{PDC}}|1\rangle\langle 1|, \\ 
		M_1^{(\mathrm{PDC})} &= \sqrt{\delta_\mathrm{PDC}} |1\rangle\langle 1|,
	\end{aligned} 
\end{equation}
where $\delta_\mathrm{PDC} \in [0, 1]$ denotes the phase-damping probability.

The ADC and PDC are typically regarded as independent and non-interacting noise mechanisms. In this work, they are modeled as a joint amplitude-phase (AP) channel, where both types of noise act simultaneously. The corresponding Kraus operator representation of this channel is written as
\begin{align}
	M_0^{(\mathrm{AP})} &= \begin{pmatrix}1 & 0 \\ 0 & \sqrt{1 - \delta_\mathrm{ADC} - \omega^\prime} \end{pmatrix} \\ \nonumber
	&= \frac{1 + \sqrt{1 - \delta_\mathrm{ADC} - \omega^\prime}}{2} \mathbb{I} + \frac{1 - \sqrt{1 - \delta_\mathrm{ADC} - \omega^\prime}}{2} \sigma_z, \\ \nonumber
	M_1^{(\mathrm{AP})} &= \begin{pmatrix}0 & \sqrt{\delta_\mathrm{ADC}} \\ 0 & 0 \end{pmatrix} = \frac{\sqrt{\delta_\mathrm{ADC}}}{2} \sigma_x + \frac{i\sqrt{\delta_\mathrm{ADC}}}{2} \sigma_y, \\ \nonumber
	M_2^{(\mathrm{AP})} &= \begin{pmatrix}0 & 0 \\ 0 & \sqrt{\omega^\prime} \end{pmatrix} = \frac{\sqrt{\omega^\prime}}{2}\mathbb{I} - \frac{\sqrt{\omega^\prime}}{2} \sigma_z, \nonumber
\end{align}
where $\omega^\prime = (1 - \delta_\mathrm{ADC}) \delta_\mathrm{PDC}$ and $\sigma_y = i|1\rangle\langle 0|-i|0\rangle\langle 1|$. Theoretically, the amplitude-damping and phase-damping probabilities are expressed as $\delta_\mathrm{ADC} = 1 - \exp(-t / T_1)$ and $\delta_\mathrm{PDC} = 1 - \exp(t / T_1 - 2t / T_2)$, respectively, where $t = 100\,\text{ns}$ is a single-qubit gate time \cite{etxezarretamartinez2021}.

\begin{table}[!t]
	\centering
	\caption{\textbf{Testing and training accuracies of AnRx-, AnRy-, and AnRz-QSVM under various physically interpretable noise backends on the known quantum attack dataset.}}
	\label{table03}
	\begin{tabular*}{\linewidth}{@{\extracolsep{\fill}}lcccccc@{}}
		\toprule
		\multirow{2}{*}{{Backend}} & \multicolumn{2}{c}{AnRx-QSVM} & \multicolumn{2}{c}{AnRy-QSVM} & \multicolumn{2}{c}{AnRz-QSVM} \\
		\cline{2-3} \cline{4-5} \cline{6-7}
		\rule{0pt}{10pt}& Test & Train & Test & Train &  Test &  Train \\
		\midrule
		LNRM & 1.0000 & 1.0000 & 0.9944 & 0.9976 & 1.0000 & 0.9976 \\
		MNSTM & 0.9889 & 0.9881 & 0.9944 & 0.9929 & 0.8778 & 0.8500 \\
		HNAM & 0.9111 & 0.8857 & 0.8722 & 0.8714 & 0.4000 & 0.3738 \\
		\bottomrule
	\end{tabular*}
\end{table}
\begin{table}[!t]
	\centering
	\caption{\textbf{Testing and training accuracies of AnRx-, AnRy-, and AnRz-QSVM under various physically interpretable noise backends on the unknown quantum attack dataset.}}
	\label{table04}
	\begin{tabular*}{\linewidth}{@{\extracolsep{\fill}}lcccccc@{}}
		\toprule
		\multirow{2}{*}{{Backend}} & \multicolumn{2}{c}{AnRx-QSVM} & \multicolumn{2}{c}{AnRy-QSVM} & \multicolumn{2}{c}{AnRz-QSVM} \\
		\cline{2-3} \cline{4-5} \cline{6-7}
		\rule{0pt}{10pt}& Test& Train & Test & Train & Test & Train \\
		\midrule
		LNRM &1.0000  &1.0000  &1.0000  &1.0000  &1.0000 & 1.0000 \\
		MNSTM &1.0000  &1.0000  &1.0000  &1.0000  &1.0000 &1.0000 \\
		HNAM &1.0000  &1.0000  &1.0000  &1.0000  &0.3667 &0.3476\\
		\bottomrule
	\end{tabular*}
\end{table}

\begin{table*}[!t]
	\centering
	\caption{\label{table05}\textbf{Comparison of AnRx-, AnRy-, and AnRz-QSVM under various physically interpretable noise backends on the testing set of the known quantum attack dataset.}}
	\begin{tabular*}{\linewidth}{@{\extracolsep{\fill}}lcccccccccc@{}}
		\toprule
		\multirow{2}{*}{{Backend}}& \multirow{2}{*}{{Model}}&
		\multicolumn{3}{c}{Macroaverage} &
		\multicolumn{3}{c}{Microaverage} &
		\multicolumn{3}{c}{Weighted average} \\
		\cline{3-5}   \cline{6-8}  \cline{9-11}  
		\rule{0pt}{10pt} &  & Precision & Recall & F1 score & Precision & Recall & F1 score & Precision & Recall & F1 score \\  
		\midrule
		LNRM
		&AnRx-QSVM  &1.0000&	1.0000&1.0000&1.0000&1.0000&1.0000&1.0000&1.0000&1.0000\\
		&AnRy-QSVM &
		0.9943 &
		0.9951 &
		0.9946 &
		0.9944 &
		0.9944 &
		0.9944 &
		0.9946 &
		0.9944 &
		0.9945 \\
		&AnRz-QSVM &
		1.0000&
		1.0000&
		1.0000&
		1.0000&
		1.0000&
		1.0000&
		1.0000&
		1.0000&
		1.0000\\
		MNSTM & AnRx-QSVM  & 0.9889    & 0.9902 & 0.9892  & 0.9889 & 0.9889 & 0.9889  & 0.9896  & 0.9889 & 0.9889 \\
		& AnRy-QSVM & 0.9943  & 0.9951 & 0.9946 & 0.9944 & 0.9944 & 0.9944   & 0.9946  & 0.9944 & 0.9945 \\
		& AnRz-QSVM & 0.9345  &0.8690  &  0.8514  &0.8778  & 0.8778 & 0.8778  & 0.9258  &0.8778  & 0.8532 \\
		HNAM &AnRx-QSVM  & 0.9467    & 0.9048 & 0.9016   & 0.9111  & 0.9111 & 0.9111   & 0.9396 & 0.9111 & 0.9018 \\
		& AnRy-QSVM & 0.9095   & 0.8641 & 0.8474   & 0.8722  & 0.8722 & 0.8722  & 0.9022  & 0.8722 & 0.8490 \\
		& AnRz-QSVM & 0.7066   &0.3761  &  0.3211  &0.4000& 0.4000 &0.4000 & 0.7008 & 0.4000 & 0.3290 \\
		\bottomrule
	\end{tabular*}
\end{table*}
\begin{table*}[!]
	\centering
	\caption{\label{table06}\textbf{Comparison of AnRx-, AnRy-, and AnRz-QSVM under various physically interpretable noise backends on the testing set of the unknown quantum attack dataset.}}
	\begin{tabular*}{\linewidth}{@{\extracolsep{\fill}}lcccccccccc@{}}
		\toprule
		\multirow{2}{*}{{Backend}}& \multirow{2}{*}{{Model}}&
		\multicolumn{3}{c}{Macroaverage} &
		\multicolumn{3}{c}{Microaverage} &
		\multicolumn{3}{c}{Weighted average} \\
		\cline{3-5}   \cline{6-8}  \cline{9-11}  
		\rule{0pt}{10pt}&  & Precision & Recall & F1 score & Precision & Recall & F1 score & Precision & Recall & F1 score \\  
		\midrule
		LNRM 
		&AnRx-QSVM  &
		1.0000&
		1.0000&
		1.0000&
		1.0000&
		1.0000&
		1.0000&
		1.0000&
		1.0000&
		1.0000\\
		&AnRy-QSVM &
		1.0000&
		1.0000&
		1.0000&
		1.0000&
		1.0000&
		1.0000&
		1.0000&
		1.0000&
		1.0000 \\
		&AnRz-QSVM &
		1.0000&
		1.0000&
		1.0000&
		1.0000&
		1.0000&
		1.0000&
		1.0000&
		1.0000&
		1.0000\\
		MNSTM & AnRx-QSVM  &1.0000   & 1.0000 & 1.0000   & 1.0000 & 1.0000 &1.0000 &1.0000 & 1.0000 & 1.0000 \\
		& AnRy-QSVM & 1.0000 &1.0000  &1.0000 &1.0000 &1.0000 &1.0000    &1.0000   & 1.0000 &1.0000  \\
		& AnRz-QSVM & 1.0000 &1.0000  &1.0000 &1.0000 &1.0000 &1.0000& 1.0000 &1.0000  &1.0000 \\
		HNAM & AnRx-QSVM  & 1.0000 & 1.0000 &1.0000 & 1.0000 &1.0000  &1.0000  & 1.0000  & 1.0000 & 1.0000 \\
		& AnRy-QSVM & 1.0000 &1.0000 & 1.0000  & 1.0000  & 1.0000 & 1.0000  & 1.0000  & 1.0000 & 1.0000 \\
		& AnRz-QSVM & 0.5356 &0.3871 & 0.3323  & 0.3667  & 0.3667 & 0.3667 & 0.5090  & 0.3667 & 0.3054 \\
		\bottomrule
	\end{tabular*}
\end{table*}

\subsubsection{Depolarizing Channel}
The DPC plays a pivotal role among the various types of hardware noise channel. For a single-qubit system, the corresponding Kraus operator representation of this channel is as follows:
\begin{equation}
	\begin{aligned}
		M_0^{(\mathrm{DPC})} &= \sqrt{1 - \delta_\mathrm{DPC}} \, \mathbb{I}, \
		M_1^{(\mathrm{DPC})} = \sqrt{\frac{\delta_\mathrm{DPC}}{3}} \, \sigma_x, \\
		M_2^{(\mathrm{DPC})} &= \sqrt{\frac{\delta_\mathrm{DPC}}{3}} \, \sigma_y, \
		M_3^{(\mathrm{DPC})} = \sqrt{\frac{\delta_\mathrm{DPC}}{3}} \, \sigma_z,
	\end{aligned}
\end{equation}
where $\delta_\mathrm{DPC}$ denotes the depolarizing probability. In DPC, the state of a qubit remains unchanged with probability $1 - \delta_\mathrm{DPC}$, while each of the Pauli operators $X$, $Y$, and $Z$ is applied with $\delta_\mathrm{DPC}/3$. Given that the PQC employed in AnR-type QSVM consist exclusively of single-qubit gates, we evaluate the impact of depolarizing noise on single-qubit gate fidelity. For a single-qubit gate affected by the depolarizing noise, the average gate fidelity is approximately given by \cite{magesan2011scalable}: $F^\prime=1-{2\delta_\mathrm{DPC}}/{3} = 1- \epsilon_1$. Therefore, the depolarizing probability takes the form $\delta_\mathrm{DPC}={3\epsilon_1}/{2}$.

As shown in Table~\ref{table02}, we construct three physically interpretable noise backends with varying noise intensities for AnR-type QSVM, based on noise metrics derived from representative superconducting quantum hardware:
\begin{itemize}
	\item \text{low\_noise\_realistic\_model} (LNRM)
	\item \text{mid\_noise\_stress\_test\_model} (MNSTM)
	\item \text{high\_noise\_adversarial\_model} (HNAM)
\end{itemize}

To realistically model gate-level errors, appropriate noise channels are inserted after each gate operation. Specifically, a depolarizing channel (denoted as $D$), an amplitude-damping channel ($A$), and a phase-damping channel ($P$) are inserted between adjacent quantum gates. Between the final gate and the measurement, we apply a depolarizing channel ($D$) and a bit-flip channel ($B$). As illustrated in Figs.~\ref{figure13}, \ref{figure14}, and \ref{figure15}, $D$, $A$, $P$, and $B$ are symbolic labels used to indicate specific noise channels rather than quantum gates.

\subsection{Benchmarking AnRx-, AnRy-, and AnRz-QSVM under Various Physically Interpretable Noise Backends}
To rigorously assess the performance of AnRx-, AnRy-, and AnRz-QSVM under various physically interpretable noise backends, we adopt a diverse set of evaluation criteria. These include overall accuracy, as well as precision, recall, and F1 score assessed under macroaverage, microaverage, and weighted average schemes \cite{grandini2020metrics}.

As shown in Table~\ref{table03}, the quantum simulation results demonstrate that noise has differential impacts on AnRx-, AnRy-, and AnRz-QSVM in the known quantum attack detection task. As the noise level increases, the detection accuracy of all three models—AnRx-, AnRy-, and AnRz-QSVM—shows a downward trend. Note that AnRz-QSVM experience the most pronounced degradation in performance. This may be attributed to their deeper quantum circuits, which involve a larger number of quantum gates and consequently introduce more noise channels during simulation, rendering them more vulnerable to cumulative noise effects. In contrast, the results from Table~\ref{table04} indicate that noise has little impact on AnRx-QSVM and AnRy-QSVM in the unknown quantum attack detection task. Only AnRz-QSVM exhibit a notable accuracy drop on the HNAM.

In addition, AnRx-QSVM demonstrate stronger robustness than the other QML variants in detecting both known and unknown quantum attacks under various physically interpretable noise backends. Interestingly, the numerical results of other evaluation criteria (see Tables~\ref{table05} and \ref{table06}) show that the accuracy is aligned with the precision, recall, and F1 score calculated using the microaverage scheme. Furthermore, they are also consistent with the recall obtained from the weighted average scheme. This consistency suggests that these evaluation criteria play equivalent roles in assessing the overall detection performance of the proposed QML-ADF, and can therefore be considered functionally equivalent to some extent.

\section{Conclusion}\label{conclusion}
To address security vulnerabilities in high-rate CV-QKD systems, we propose QML-ADF, a QML-based framework for noise-resistant and feature-aware attack detection. By systematically designing and benchmarking variants of QNN and QSVM, the proposed QML-ADF effectively detects both known and previously unknown quantum attacks. The AnR-type QSVM achieve the highest accuracy, reaching 100\% in attack detection under ideal conditions. Robustness is further validated on three physically interpretable noise backends, constructed from noise metrics of real superconducting quantum hardware. Remarkably, the AnRx-QSVM maintain robust detection performance even under the HNAM with strong noise, showing less than a 10\% drop in accuracy for known quantum attacks and no degradation for unknown quantum attacks.

These findings underscore the potential of QML in enhancing the security of next-generation high-rate quantum communication infrastructures and open new avenues for research in quantum cryptography and secure quantum communications. Building on this work, future efforts may extend the proposed framework to quantum communication protocols beyond CV-QKD.

\ifCLASSOPTIONcompsoc
\section*{Acknowledgments}
\else
\section*{Acknowledgment}
\fi

This work was supported by the A*STAR under Grant Nos.~M21K2c0116 and M24M8b0004; the Singapore National Research Foundation under Grant Nos.~NRF-CRP22-2019-0004, NRF-CRP30-2023-0003, NRF-CRP31-0001, NRF2023-ITC004-001, and NRF-MSG-2023-0002; and the Singapore Ministry of Education Tier 2 under Grant Nos.~MOE-T2EP50221-0005 and MOE-T2EP50222-0018. This work was also supported by the National Natural Science Foundation of China under Grant No.~62076091 and the Natural Science Foundation of Hunan Province under Grant No.~2023JJ30167. Chao Ding is supported by the China Scholarship Council under Grant No.~202306130112. Professor Weibo Gao is the Dieter Schwarz Endowed Professor in QUASAR at Nanyang Technological University, supported by the Dieter Schwarz Stiftung gGmbH.



\begin{IEEEbiography}[{\includegraphics[width=1in,height=1.25in,clip,keepaspectratio]{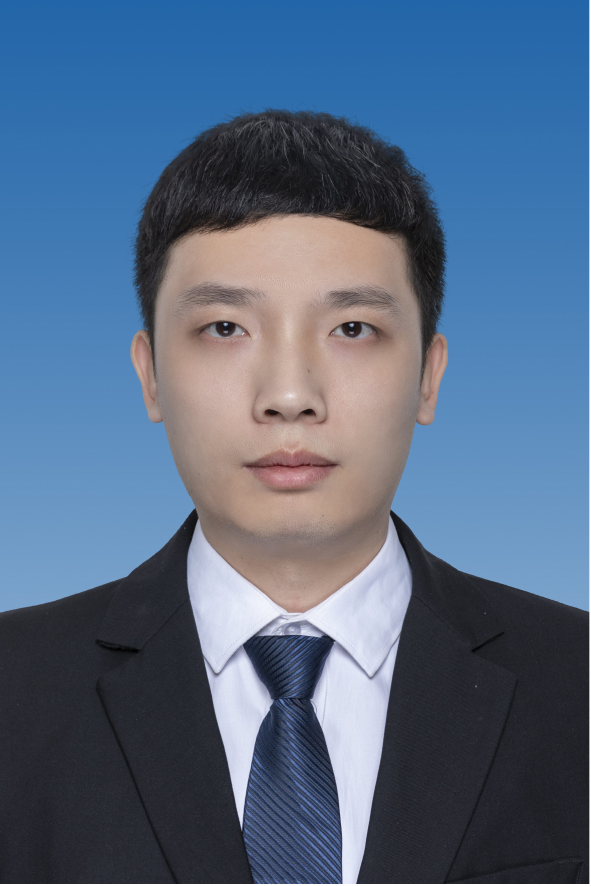}}]
	{Chao Ding} received the M.S. degree in software engineering from the Central South University, Changsha, China, in 2021. He is currently pursuing a Ph.D. degree with the National Engineering Research Center for Robot Visual Perception and Control, Hunan University, Changsha, China. He was also a visiting Ph.D. student at the School of Physical and Mathematical Sciences (SPMS), Nanyang Technological University (NTU). He is currently interning at the Centre for Quantum Technologies (CQT), National University of Singapore (NUS). His research interests include quantum machine learning and quantum communication.
\end{IEEEbiography}
\begin{IEEEbiography}[{\includegraphics[width=1in,height=1.25in,clip,keepaspectratio]{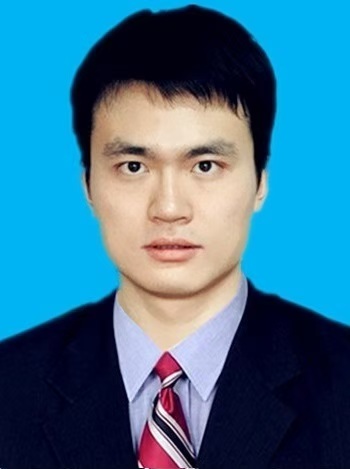}}]
	{Shi Wang} received the Ph.D. degree in engineering and computer science from the Australian National University, Canberra, Australia, in 2014. From 2013 to 2014, he was a Postdoctoral Fellow with the National Institute of Informatics, Tokyo, Japan. He is currently an Associate Professor with the College of Electrical and Information Engineering, Hunan University, Changsha, China. His research interests include quantum coherent feedback control, quantum network analysis and synthesis, quantum machine learning, and multiagent systems.
\end{IEEEbiography}
\begin{IEEEbiography}[{\includegraphics[width=1in,height=1.25in,clip,keepaspectratio]{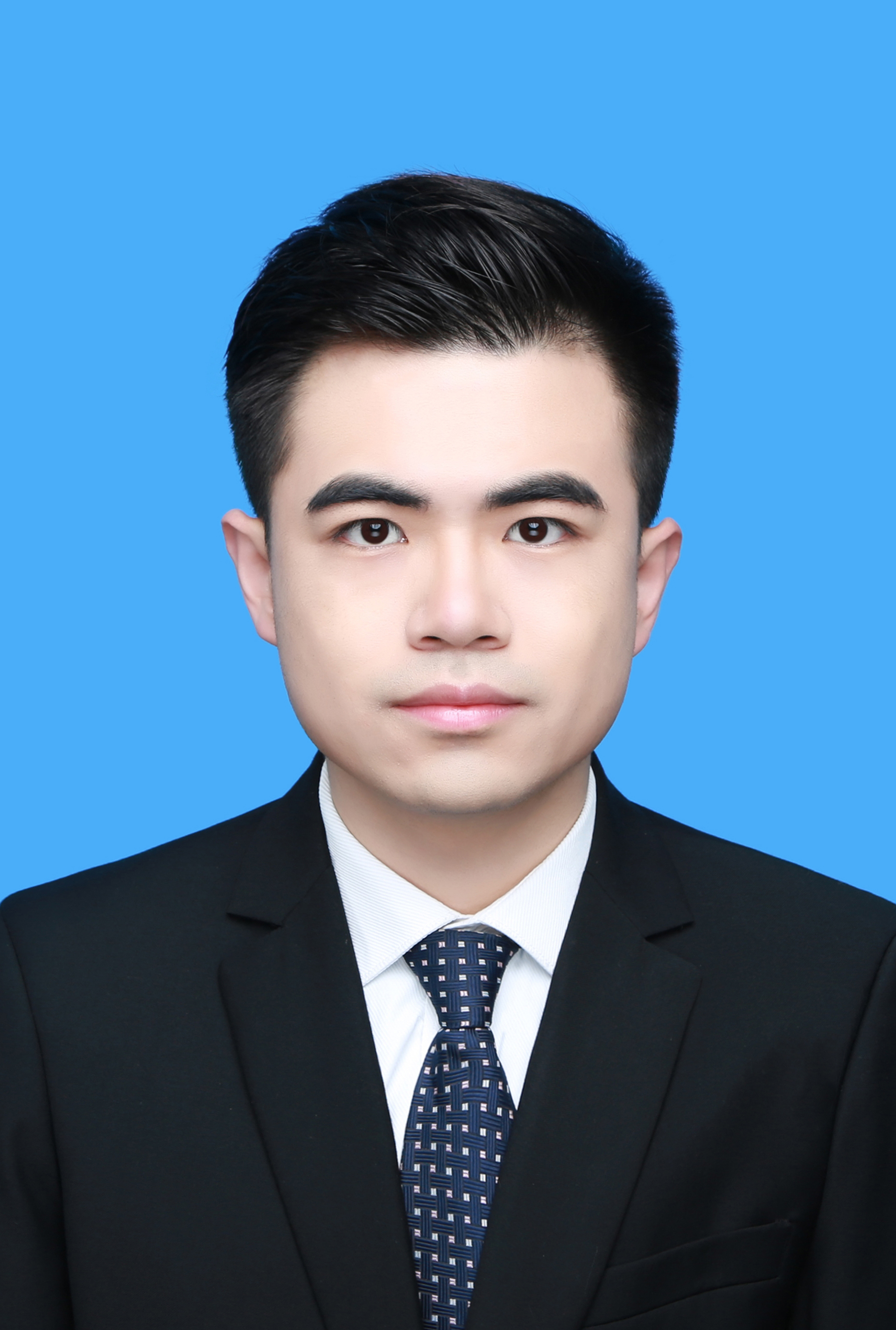}}]
	{Jingtao Sun} received the B.S. and M.S. degree in the College of Electrical and Information Engineering from Hunan University, Changsha, China, where he is currently working toward the Ph.D. degree with the National Engineering Research Center for Robot Visual Perception and Control, Hunan University. He was also a visiting Ph.D student at the Department of Electrical and Computer Engineering (ECE), National University of Singapore (NUS). His research interests include 3D computer vision, robotics, and multi-modal.
\end{IEEEbiography}
\begin{IEEEbiography}[{\includegraphics[width=1in,height=1.25in,clip,keepaspectratio]{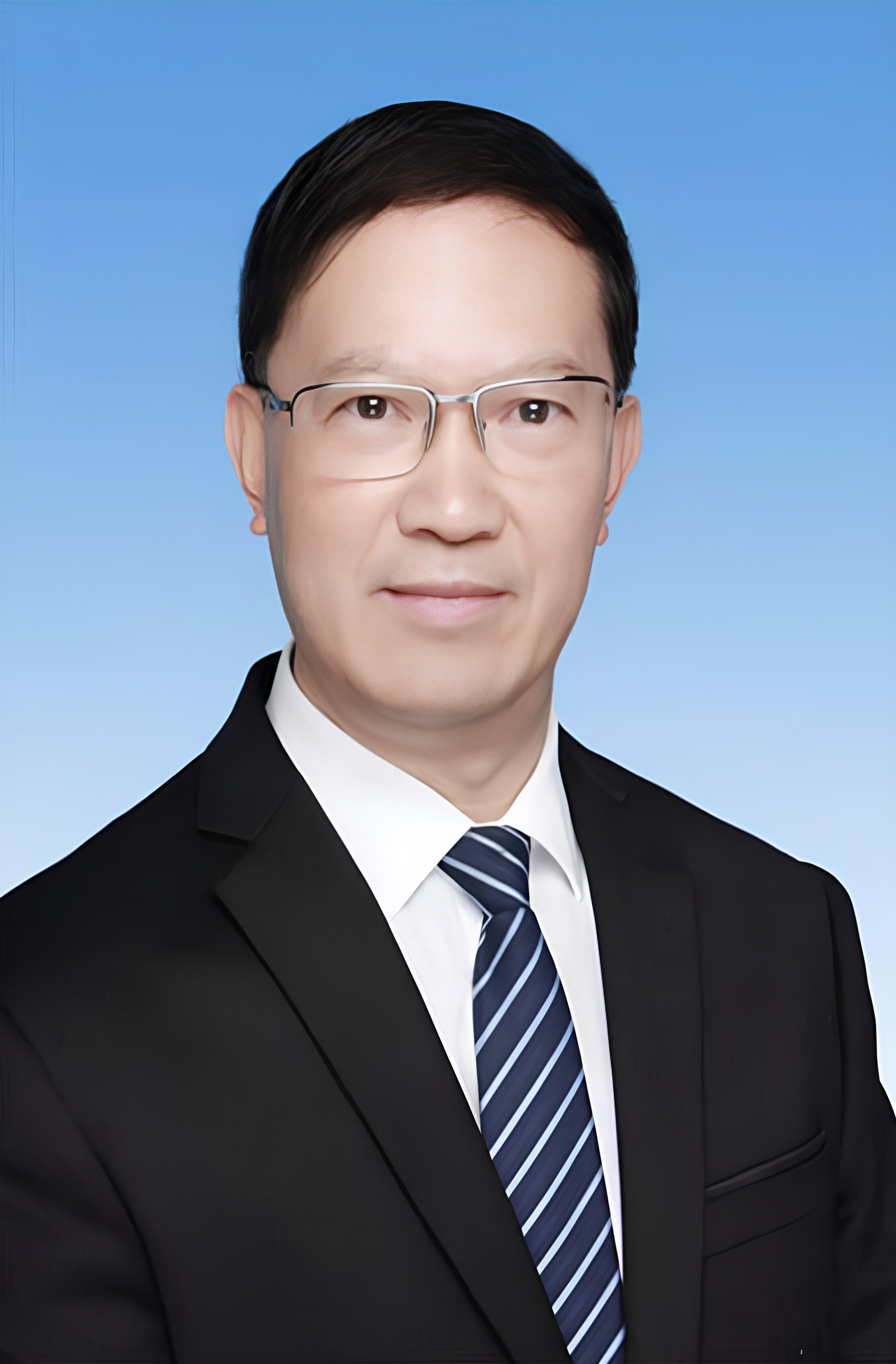}}]
	{Yaonan Wang} received the Ph.D. degree in electrical engineering from Hunan University, Changsha, China, in 1994. He was a Postdoctoral Research Fellow with the Normal University of Defense Technology, Changsha, from 1994 to 1995. From 1998 to 2000, he was a Senior Humboldt Fellow in Germany and, from 2001 to 2004, was a visiting Professor with the University of Bremen, Bremen, Germany. Since 1995, he has been a Professor with the College of Electrical and Information Engineering, Hunan University. He is an Academician with the Chinese Academy of Engineering.
\end{IEEEbiography}
\begin{IEEEbiography}[{\includegraphics[width=1in,height=1.25in,clip,keepaspectratio]{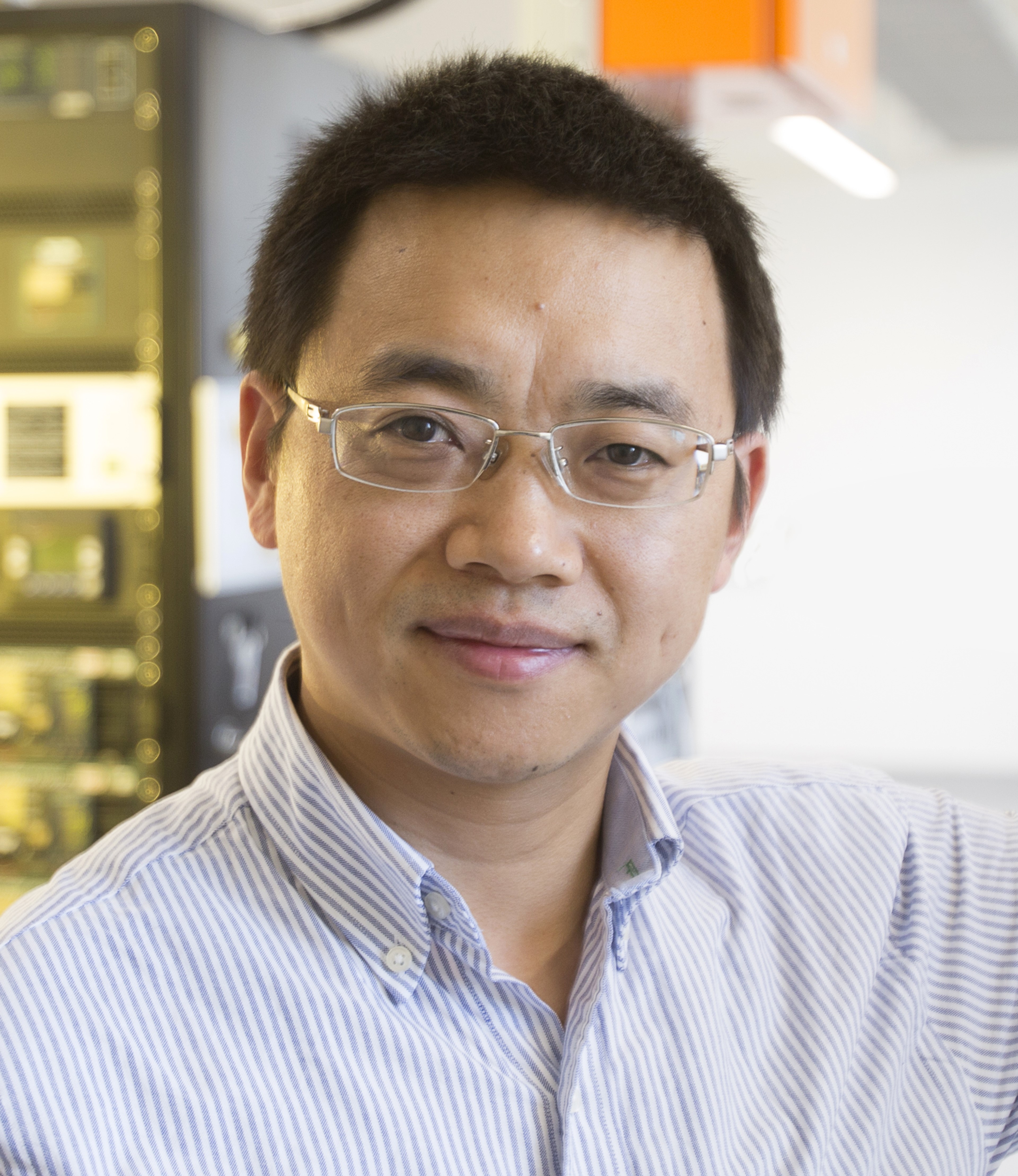}}]
	{Daoyi Dong}(Fellow, IEEE) is currently a Professor and an ARC Future Fellow at the Australian Artificial Intelligence Institute, University of Technology Sydney, Australia and an Honorary Professor at the Australian National University. He was with the Australian National University, the University of New South Wales, Australia, the Institute of Systems Science, Chinese Academy of Sciences and Zhejiang University.
	
	His research interests include quantum control, quantum estimation and machine learning. Prof. Dong was awarded an ACA Temasek Young Educator Award by The Asian Control Association and is a recipient of a Future Fellowship, an International Collaboration Award and an Australian Post-Doctoral Fellowship from the Australian Research Council, and a Humboldt Research Fellowship from the Alexander von Humboldt Foundation of Germany. He is a Vice President of IEEE Systems, Man and Cybernetics Society, and a member of Board of Governors, IEEE Control Systems Society. He is currently an Associate Editor of Automatica and IEEE Transactions on Cybernetics. He is a Fellow of the IEEE, and a Fellow of the Australian Institute of Physics.
\end{IEEEbiography}
\begin{IEEEbiography}[{\includegraphics[width=1in,height=1.25in,clip,keepaspectratio]{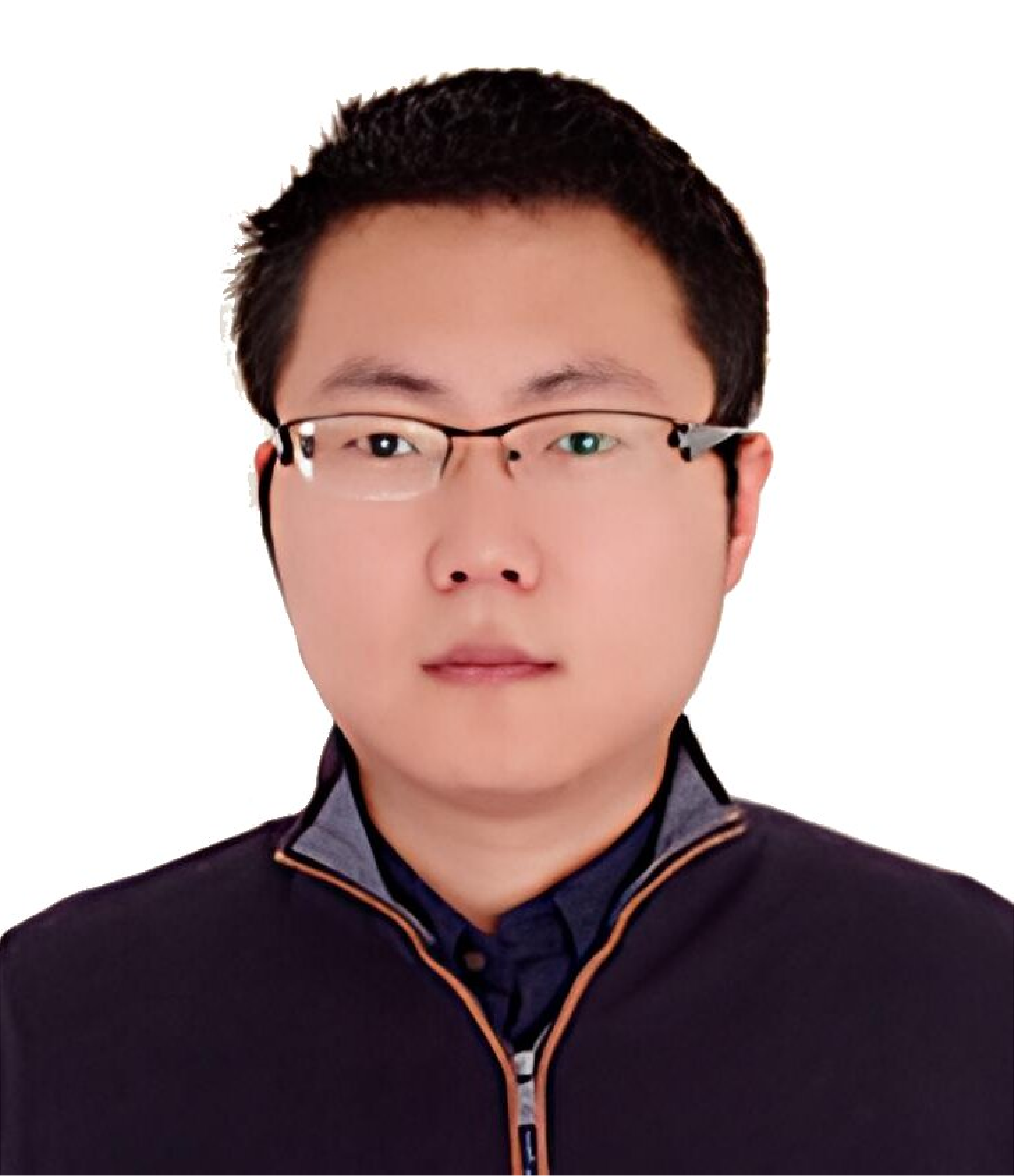}}]
	{Weibo Gao} is currently a Professor and Dieter Schwarz Endowed Professor in Quantum Sovereignty and Resilience (QUASAR) at Nanyang Technological University (NTU), Singapore. He serves as Chair of the School of Electrical and Electronic Engineering (EEE) and Director of the Centre for Quantum Technologies at NTU (CQT@NTU). He is also affiliated with the Centre for Quantum Technologies at the National University of Singapore (CQT@NUS). His research interests include quantum information and quantum optics.
	
	Prof. Gao was awarded several prestigious honors, including the Singapore President’s Young Scientist Award (YSA), the Innovators Under 35 – EmTech Asia by MIT Technology Review, and the National 100 Excellent Doctoral Dissertation Award (China). He served as an Associate Editor for Photonics Research. He currently serves on the Executive Editorial Board of Materials for Quantum Technology and the Editorial Board of Chinese Physics B.
\end{IEEEbiography}


\ifCLASSOPTIONcaptionsoff
  \newpage
\fi

\end{document}